\documentclass[a4paper,12pt]{article}

\usepackage{typearea}
\usepackage{t1enc}
\usepackage{mathrsfs}
\usepackage{graphicx}
\usepackage{multirow}
\usepackage{hyperref}
\usepackage{multirow}
\usepackage{bbm}
\usepackage{amsmath}
\usepackage{arydshln}
\usepackage{tikz}
\usepackage{authblk}
\usepackage[hang,small,bf]{caption}
\usepackage{ulem}

\usepackage[utf8]{inputenc}

\newcommand{\qtilde}{\tilde{q}}

\usepackage{tikz}
\usetikzlibrary{shadows}
\usetikzlibrary{arrows,shapes,positioning}
\usetikzlibrary{decorations.markings}
\usetikzlibrary{decorations.pathmorphing}
\usetikzlibrary{decorations.pathreplacing}
\tikzstyle arrowstyle=[scale=1]
\tikzstyle directed=[postaction={decorate,decoration={markings, mark=at position .65 with {\arrow[arrowstyle]{stealth}}}}]
\tikzstyle end directed=[postaction={decorate,decoration={markings, mark=at position 1 with {\arrow[arrowstyle]{stealth}}}}]
\tikzstyle reverse directed=[postaction={decorate,decoration={markings, mark=at position .65 with {\arrowreversed[arrowstyle]{stealth};}}}]
\tikzstyle{ann} = [fill=white,font=\footnotesize,inner sep=1pt]

\title{Updated lattice results for parton distributions}

\author[a,b]{\normalsize Constantia Alexandrou}
\author[c,d]{Krzysztof Cichy}
\author[e]{Martha Constantinou}
\author[a]{Kyriakos Hadjiyiannakou}
\author[f]{Karl Jansen}
\author[f]{Fernanda Steffens}
\author[f]{Christian Wiese}

\affil[a]{\footnotesize Department of Physics, University of Cyprus, P.O. Box 20537, 1678 Nicosia, Cyprus}
\affil[b]{The Cyprus Institute, 20 Kavafi Str., Nicosia 2121, Cyprus}
\affil[c]{Goethe-Universit\"at Frankfurt am Main, Institut f\"ur Theoretische
Physik, Max-von-Laue-Strasse 1, 60438 Frankfurt am Main, Germany}
\affil[d]{Faculty of Physics, Adam Mickiewicz University, Umultowska 85, 61-614 Pozna\'{n}, Poland}
\affil[e]{Temple University, 1925 N. 12th Street, Philadelphia, PA 19122, USA}
\affil[f]{John von Neumann Institute for Computing (NIC), DESY, Platanenallee 6, 15738 Zeuthen, Germany}

\date{\small July 11, 2017}

\begin{document}
\maketitle

\begin{abstract}
We provide an analysis of the $x$-dependence of the bare
unpolarized, helicity and transversity iso-vector parton distribution
functions (PDFs) from lattice calculations employing
(maximally) twisted mass fermions.
The $x$-dependence of the calculated PDFs resembles the one of the phenomenological
parameterizations, a feature that makes this approach very promising. 
Furthermore, we apply momentum smearing for the
relevant matrix elements to compute the lattice PDFs and
find a large improvement factor when compared to
conventional Gaussian smearing. This allows us to extend the lattice computation
of the distributions to higher values of the nucleon momentum,which is essential for the prospects of a reliable extraction of the PDFs in the future.
\end{abstract}

\section{Introduction}

Finding a way to a computation of parton distribution
functions (PDFs) with lattice QCD techniques has been a long
standing goal in lattice gauge theory.  The basic problem
roots in the Euclidean nature of lattice QCD that does not
allow for a direct calculation of the PDFs, which are usually
defined as light cone correlations in the rest frame of the target.
Therefore, lattice QCD computations have focused on
moments of PDFs, form factors and related quantities.  The
calculations have been  very successful with important
results connecting to phenomenology and experiment,  see
\cite{Constantinou:2014tga, Constantinou:2015agp,
Alexandrou:2015yqa, Alexandrou:2015xts, Syritsyn:2014saa}
for recent reviews and \cite{Lin:2015dga} for an overview of
lattice activities on hadron structure.

Moreover, lattice QCD calculations concerned with hadron
structure are now being performed at or close to the
physical value of the pion mass \cite{Abdel-Rehim:2015owa,
  Bali:2016wqg, Bhattacharya:2015wna, Durr:2015dna,
Ohta:2014rfa, Green:2014xba, Green:2012ud, Lin:2014gaa}
allowing for a significantly improved control of the
involved systematic uncertainties, since the previously
required chiral extrapolation to the physical value of the
quark mass can be avoided.  In addition, in recent works
also the disconnected diagrams --which were often neglected
in the past -- have been taken into account, involving a very
high statistics of $\mathcal{O}(10^5)$ measurements
\cite{Alexandrou:2013wca,Abdel-Rehim:2013wlz, Chambers:2015bka, Bhattacharya:2015esa}.

Despite this success, a direct calculation of PDFs remains
highly desirable for several reasons. First, having the
non-perturbative functional shape of the PDFs over a broad
range of the momentum fraction (Bjorken variable) $x$
available would provide essential information on the
structure of hadrons, as predicted solely by QCD. Furthermore,
the knowledge of the PDFs would allow for a direct
comparison to experimental results and  phenomenological
analyses of deep inelastic scattering data.
In particular, the flavor structure of the nucleon sea is
highly nontrivial, and the observed asymmetry between the
up and down antiquark distributions (see \cite{Chang:2014jba} for a review) is
an intrinsically nonperturbative QCD effect \cite{Thomas:1983fh,Thomas:2000ny}.
However, the lattice calculations have been restricted to
the first two or three moments of PDFs, and thus
providing only limited insight into the structure of hadrons.

A possible way towards a direct calculation of PDFs has been
proposed in Ref.~\cite{Ji:2013dva}. The idea is to
compute the so-called quasi distributions where the Wilson line
connecting the quarks in the nucleon is taken in the
spatial directions, avoiding thus the difficulty of the light
cone dominance.  The quasi distributions can be related to
the PDFs through a suitable matching procedure. In the language of
an effective field theory, it means that parton distributions
can be extracted from the lattice observables using a systematic expansion in
the inverse powers of the nucleon momentum \cite{Ji:2014gla}.
The approach of
Ref.~\cite{Ji:2013dva} has already been tested for the
bare distribution functions in
\cite{Lin:2014zya,Chen:2016utp,Alexandrou:2015rja,Gamberg:2014zwa} and it
could be demonstrated that at least on a qualitative level,
the shape of the physical distribution functions is
reproduced.  A most remarkable finding of these lattice
calculations has been that the quark-antiquark asymmetry in
the PDFs does come out automatically from the first principles
lattice QCD calculations without any additional input.

In this paper, we extend the
computation of PDFs on the lattice, compared to our previous
lattice study \cite{Alexandrou:2015rja}, in several ways:
in addition to the unpolarized PDFs,
we now also include the helicity and the transversity PDFs.
Also, all results presented in this paper (using the standard Gaussian smearing of quark fields) have a
substantially increased statistics of about 30000
measurements, which is about a factor of 6 improvement
compared to \cite{Alexandrou:2015rja}.  This allows us to go
to larger momenta and thus a much better control of the
matching to the physical distribution can be achieved.  We
also provide a test of the recent new matching formula of
Ref.~\cite{Chen:2016fxx}.  As a new, more technical step, we
have implemented a recently developed type of smearing of quark fields, the momentum smearing \cite{Bali:2016lva},
and we find a significant improvement of
the signal-to-noise ratio when compared to the previous
momentum independent (Gaussian) smearing. Finally, for illustration purposes, we
demonstrate how the method of quasi distributions works in
practice, by performing the necessary steps of this
procedure at tree-level of perturbation theory, i.e. for
free quarks and show how the expected $\delta$-function
distribution is approached.

The still open question, which however goes much beyond the
scope of this paper, is the renormalization of the matrix
elements needed to compute the quasi distributions. Although
there are already first works towards the renormalization of
PDFs \cite{Ji:2015jwa,Ishikawa:2016znu,Chen:2016fxx}, a
full, in particular  non-perturbative analysis, including a
subtraction of power-like divergences combined with
non-perturbative lattice calculations is still missing.
However, work in this direction is in progress by us and
will be discussed in the near future in a separate work\footnote{After the submission of this manuscript we have indeed developed a full renormalization prescription~\cite{Constantinou:2017sej,Alexandrou:2017huk}.}.

The paper is organized as follows. In Section 2, we
provide the theoretical principles and the setup to compute the quasi distributions on the
lattice. In Section 3, numerical results are presented, including
the matching to physical quark distributions. We finally
conclude and discuss the prospects of this approach in Section 4.

\section{Theoretical principles and lattice techniques}

\subsection{Quasi-parton distributions and matching to physical parton distributions}

Parton distributions can be defined either in the infinite
momentum frame (IMF), according to Feynman's original
proposal \cite{Feynman:102074}, or in the rest frame of the
nucleon, this last one being the definition usually found in
the literature. In the original definition, because the
nucleon is with infinite momentum, there is no time for the
partons to interact and thus they are
essentially free. On the other hand, in the nucleon rest
frame, the distributions are given by light cone
correlations 
\begin{equation}
q(x,\mu) = \int_{-\infty}^{+\infty}\frac{d\xi^-}{4\pi}e^{-ixP^+\xi^-}\langle P|\overline{\psi}(\xi^-)\gamma^+W(\xi^-,0)\psi(0)|P\rangle ,
\label{eq1}
\end{equation}
where $\gamma^+$ is the Dirac structure in the unpolarized case, $\mu$ is the renormalization scale, $\xi^-=(\xi^0 - \xi^3)/\sqrt{2}$,
$P^+=(P^0+P^3)/\sqrt{2}=M/\sqrt{2}$, and
$W(\xi^-,0)=e^{-ig\int_0^{\xi^-}d\eta^- A^+(\eta^-)}$ is the
Wilson line connecting the point $0$ to the point $\xi^-$.
This definition is completely equivalent to that of the IMF
\cite{Jaffe:1985je} meaning that, in principle, one could
calculate the distributions using any of the two approaches.
In practice, however, if we use lattice QCD, we simply cannot
have infinite momentum because the maximum momentum that can be reached
on the lattice is limited by the finite lattice spacing $a$.
We are thus left with Eq.\,(\ref{eq1}), which, in turn,
cannot be calculated on the lattice, because it is light cone
dominated, $\xi^2 = t^2 - \vec{r}^2 \sim 0$, i.e.\ we
have access to a single point only in Euclidean space. Nevertheless, we can
calculate the quasi distributions, which are defined (in the unpolarized case) as
\begin{equation}
\label{eq2}
\qtilde (x,  P_3) = \int_{-\infty}^\infty \frac{dz}{4\pi} e^{-izk_3} \langle P |
\bar{\psi}(0,z)\gamma^3 W(z) \psi(0,0) |P\rangle
+ \mathcal{O}\left(\frac{\Lambda_{QCD}^2}{P_3^2},\frac{M^2}{P_3^2}\right),
\end{equation}
where $P=(M,0,0,P_3)$,  $k_3 = x P_3$ is the quark momentum
in the $z$-direction, and $W(z) = e^{-ig \int_0^z dz^{'}
A_3(z^{'})}$. Because of the finite momentum, some quarks
can carry more momentum than the nucleon itself or even move
backwards, and because of this one can have $x>1$ or $x<0$,
and the usual partonic interpretation is lost. Also,
there are higher twist (HT) and target mass corrections
(TMCs) that need to be applied. In particular, the TMCs are
essential to bring the quasi distributions to
their correct support in $x$ \cite{Alexandrou:2015rja}.

The quasi distributions represent the would-be distributions
of a nucleon moving in one particular direction, the third
direction as defined in Eq.\,(\ref{eq2}), with a large but
finite momentum. Because the infrared physics is the same
for a nucleon with infinite momentum and a nucleon moving
with a finite (and large) momentum, the difference between
the quark distributions and the quark quasi distributions
should be in the ultraviolet region (UV) only, and thus can
be perturbatively calculated \cite{Xiong:2013bka, Alexandrou:2015rja},
\begin{eqnarray}
\label{invq}
q(x, \mu) &=& \tilde{q} (x,\Lambda,P_3) - \frac{\alpha_s}{2\pi} \tilde{q} (x,\Lambda,P_3)  \delta Z^{(1)} \left( \frac{\mu}{P_3}, \frac{\Lambda}{P_3}\right)  \nonumber\\*
& & - \frac{\alpha_s}{2\pi} \int_{-x_c}^{-|x|/ x_c} Z^{(1)} \left( \xi, \frac{\mu}{P_3}, \frac{\Lambda}{P_3} \right) \tilde{q} \left( \frac{x}{\xi} ,\Lambda,P_3 \right) \frac{d\xi}{|\xi |} \nonumber\\*
& & - \frac{\alpha_s}{2\pi} \int_{+|x|/ x_c}^{+x_c} Z^{(1)} \left( \xi, \frac{\mu}{P_3}, \frac{\Lambda}{P_3} \right) \tilde{q} \left( \frac{x}{\xi} ,\Lambda,P_3 \right) \frac{d\xi}{|\xi |} +  \mathcal{O}(\alpha_s^2).
\end{eqnarray}
Here, $\Lambda$ is the UV cutoff and $x_c \sim \Lambda/P_3$ is the maximum
$x$ value for a nonzero $\tilde{q}(x,\Lambda,P_3)$.
For the wave function, $\delta Z^{(1)}$, and the vertex,
$Z^{(1)}$, corrections, we employ the recent results of \cite{Chen:2016fxx}.
In this work, the linearly divergent terms in $\Lambda/P_3$, present
in $\delta Z^{(1)}$ and in $Z^{(1)}$, are removed through the addition of
a mass counterterm, implying that the only divergence remaining in the
matching is logarithmic (see, for instance, Eq.\,(A9) of \cite{Alexandrou:2015rja}).
The linear divergence appearing in the lattice calculation of
$\tilde{q} \left( x ,\Lambda,P_3 \right)$, however, remains. More on this
point can be found in Section 2.2.

In the end,
we want the momentum to be as large as possible,
so that any correction dependent on the finite value of $P_3$ is sufficiently small,
and the matching between the quasi distributions and the
distributions, encapsulated by Eq.\,(\ref{invq}), is valid.
In the next subsection, we will discuss how this can be achieved on the lattice.

In this work, we calculate matrix elements of operators
with the Dirac structure in Eq.~(\ref{eq2}):
\begin{itemize}
  \item $\gamma_3$, for the case of the
    unpolarized quasi distributions $\tilde{q}(x,\Lambda,P_3)$\,,
  \item $\gamma_3 \gamma_5$, for the case of the helicity
    quasi distributions $\Delta \tilde{q}(x,\Lambda,P_3)$\,,
  \item $\gamma_3 \gamma_j$ ($j=1,2$), for the case of the transversity
  quasi distributions $\delta \tilde{q}(x,\Lambda,P_3)$\,.
  \end{itemize}

\subsection{Matrix elements}
The relation between the quasi distributions
and the matrix elements for the unpolarized case is
\begin{equation}
\label{MElements}
\langle P |\bar{\psi}(0,z)\gamma_3 W(z) \psi(0,0) |P\rangle
= \overline{u}(P)h(P_3,z)u(P)\,,
\end{equation}
with
\begin{equation}
\label{QuasiDistributions}
\tilde{q}(x,\Lambda,P_3)=2P_3
\int_{-\frac{L}{2}}^{+\frac{L}{2}}
\frac{dz}{4\pi}e^{-ixP_3z}h(P_3,z)\,,
\end{equation}
where $\Lambda=1/a$, $a$ is the lattice spacing and $L$ is the spatial extent of the lattice. For the helicity and transversity
quasi distributions, one only replaces $\gamma_3$ by
the desired Dirac structure.

The required matrix elements are obtained from the ratio of
suitable two- and three-point functions. The three-point
function is constructed in the usual way with boosted
nucleon interpolating fields and a local operator,
\begin{equation}
C^{{3pt}}(t,\tau,0;\vec{P}) = \left \langle N_{\alpha}(\vec{P},t) \mathcal O(\tau) \overline{N}_{\alpha}(\vec{P},0)\right \rangle,
\label{C3pt}
\end{equation}
where $\langle ... \rangle$ is the average over a
sufficiently large number of gauge field configurations. The
boosted nucleon field is defined via the Fourier
transformation of quark fields in position space,
\begin{equation}
\label{NucleonInterpolation}
N_{\alpha}(\vec{P},t) = \Gamma_{\alpha\beta} \sum_{\vec{x}}{e}^{-i \vec{P} \vec{x}}\epsilon^{abc}u_{\beta}^a(x)\left( {d^b}^T(x)\mathcal C \gamma_5 u^c(x)\right),
\end{equation}
where $\mathcal C$ is the charge conjugation matrix, chosen to be $i\gamma_0\gamma_2$, and
$\Gamma_{\alpha\beta}$ is the parity projector, depending on
the Dirac structure used.  We use here the parity plus
projector $\Gamma = \frac{1+\gamma_4}{2}$ for the case of
$\gamma_3$,  $\Gamma = i\gamma_3 \gamma_5
\frac{1+\gamma_4}{2}$ for the case of $\gamma_3 \gamma_5$ ,
and  $\Gamma = i \gamma_k \frac{1+\gamma_4}{2}$ (with $k
\neq j \neq 3$) for the case of $\gamma_3 \gamma_j$.  The
operator at vanishing momentum transfer ($Q^2=0$) to be
inserted in Eq.\,(\ref{C3pt}) is obtained by choosing
\begin{equation}
\label{EuclideanOperator}
\mathcal O(z, \tau, Q^2=0) = \sum_{\vec{y}}\overline{\psi}(y
+ \hat{e}_3 z)\gamma_3 W_3(y+ \hat{e}_3 z,y)\psi(y),
\end{equation}
for the case of $\gamma_3$, where $y=(\vec{y},\tau)$.
Similar expressions hold for the cases of  $\gamma_3
\gamma_5$ and $\gamma_3 \gamma_j$. The Wilson line is
computed as a product of gauge links along the chosen axis,
where only the shortest path is considered
\begin{equation}
W_j(y+z\hat{e}_j,y)=U_j(y+(z-1)\hat{e}_j)\dots U_j(y+\hat{e}_j)U_j(y).
\label{WilsonLine}
\end{equation}
Due to the rotational invariance on the lattice, we are
certainly not restricted to the 3 direction and can easily
generalize the above expressions to the other two
directions.

To complete the calculation, we also need the two-point
function, which is also constructed from the nucleon
interpolating field as in Eq.\,(\ref{C3pt}), but without the
insertion of the operator. With this in mind, the desired
matrix element for the case of $\gamma_3$ is extracted from
\begin{equation}
\label{Ratio3to2}
\frac{C^{{3pt}}(t,\tau,0;\vec{P})}{C^{{2pt}}(t,0;\vec{P})}\stackrel{0\ll \tau\ll t}{=}\frac{-iP_3}{E}h(P_3,\Delta z),
\end{equation}
with $E=\sqrt{(P_3)^2+M^2}$ the total energy of the nucleon.
For the helicity, $\Delta h(P_3,\Delta z)$, and the
transversity, $\delta h(P_3,\Delta z)$, matrix elements, the
pre-factor $\frac{-iP_3}{E}$ is absent, which can be easily
verified from their definition from Eq.\,(\ref{MElements}).

Due to the symmetric structure of the operator, there is a
relation between the negative and positive $z$ direction of
the matrix elements. To see this, one can apply the gauge
link identity $U_{-j}(x) = U_{j}(x -
\hat{e}_j)^\dagger$ to Eq.\,(\ref{WilsonLine}), which
results in
$W_j(y+z\hat{e}_j,y)=W_j(y,y+z\hat{e}_j)^\dagger$. Thus, one
obtains the relation $\mathcal O(z, \tau)=-\mathcal O(-z,
\tau)^\dagger$ for the Euclidean operator defined in
Eq. (\ref{EuclideanOperator}), due to its translational
invariance structure. For the cases of $\gamma_3 \gamma_5$
and $\gamma_3 \gamma_j$, however, one has $\mathcal O(z,
\tau)=+\mathcal O(-z, \tau)^\dagger$. Taking into account
the pre-factors of Eq.\,(\ref{Ratio3to2}), one then obtains
\begin{eqnarray}
\label{Asymmetry}
h(P_3,z)&=&h(P_3,-z)^\dagger , \nonumber\\*
\Delta h(P_3,z)&=&\Delta h(P_3,-z)^\dagger , \nonumber\\*
\delta h(P_3,z)&=&\delta h(P_3,-z)^\dagger .
\end{eqnarray}
These are completely general equations that not only can be
used as a cross-check to our lattice results, but also have
the fundamental consequence of producing an asymmetry
between the quark and antiquark distributions, as it will
be clear in Section \ref{SEC_QD}. For the operators themselves, we will
compute only the iso-vector quark combination, i.e.\ a
$\tau^3$ matrix in flavor space is inserted, as this avoids
possible operator mixing and disconnected contributions.
Consequently, the resulting matrix elements will carry a $u-d$
superscript.

\subsection{Smearing of quark fields}
Usually the signal-to-noise ratio for a boosted nucleon is
not strong, and to enhance it one introduces Gaussian
smearing ($S$) in the quark fields
\cite{Alexandrou:1992ti,Gusken:1989qx}. The results
presented in \cite{Alexandrou:2015rja, Alexandrou:2016bud}
use such smearing and one sees that the errors increase
rapidly with the injected momentum, making this
approach unfeasible for $P_3 > 6\pi/L$. This upper limit
imposes a serious constraint on the simulations, because we
want to reach values for the momentum where higher order
corrections in $\alpha_s$ to Eq.\,(\ref{invq}) can be safely
neglected. Also, we want the corrections from the matching
itself, as well as the ones from TMCs and from HTs, to be
small. In other words, it is desirable to have a nucleon
with momentum that, in practice, is large enough such that the
lattice data start to show saturation, in the sense that
the computed quasi distributions are essentially unchanged
as the momentum grows.

In Ref.~\cite{Bali:2016lva}, a new type of smearing, called
momentum smearing ($S_{mom}$), was proposed where the quark
fields are modified as
\begin{equation}
S_{mom}\psi(x) = \frac{1}{1+6\alpha}(\psi(x) + \alpha \sum_j U_j(x)e^{ik\hat{e}_j}\psi(x+\hat{e}_j)),
\label{momsmearing}
\end{equation}
where $k=\zeta P$, with $P$ the lattice momentum of the
nucleon and $\zeta$ a tunable parameter, $\alpha$ is a
positive constant, and $U_j(x)$ are the gauge links in the
direction $j$. Throughout this work, we use 50 steps of smearing
with $\alpha=4$. As for $\zeta$, we use 0.45, which is the value
found to be optimal in \cite{Bali:2016lva}.
The usual Gaussian smearing $S\psi(x)$
corresponds to dropping the exponential factor in
(\ref{momsmearing}). The immediate consequence of the
exponential factor is that, when going to momentum space,
the smearing of the field is not around 0, but around the
shifted momentum $k$. This has the enormous advantage that
we can tune the smearing to be done for higher values of the
nucleon momentum, and not only to values close to zero, as
with the standard Gaussian smearing, implying that the quark field smearing may now
be effective for large values of the boosted nucleon
momentum. In the next section, we explore this possibility numerically.

When computing the three-point functions using Gaussian
smearing, we have some
freedom on how to treat the quark propagator connecting the
sink position with the point where the operator is inserted.
In principle, there are two different ways of computing
this all-to-all propagator.  The first one is to use the
sequential method \cite{Martinelli:1988rr}. The shortcoming
of this method is that the sink position and the nucleon
momentum at the sink have to be fixed, making this approach
not optimal if we want to compute the matrix elements at
several values of the nucleon momentum.

The second method is the stochastic method
\cite{Alexandrou:2013xon} that uses stochastic $Z^4$
noise sources on a single time slice for the
computation of the all-to-all propagator. This method is
very flexible when using different momenta and projectors,
however it adds stochastic noise to the calculation.

In our previous work \cite{Alexandrou:2015rja,
Alexandrou:2014pna}, we tested both methods and found that at
the expense of the same computation effort, they were
compatible. Nonetheless, the stochastic method is more flexible
when computing several values of the boosted nucleon
momentum, thus our initial choice was to use stochastic
sources, even if this introduces more noise to the
system.

However, due to the recent introduction of momentum smearing
(cf.\ Eq.\,(\ref{momsmearing})), the premises that led to the
choice of the stochastic method are not satisfied anymore. When
using momentum smearing, one reduces the noise for larger
momenta dramatically, with the drawback of having to perform
separate inversions for each momentum. Thus, the flexibility
of the stochastic method is lost and it is preferable to use
the sequential method. In this way, in the present work, we use
the stochastic method for the standard Gaussian smearing and the sequential method when
using the new momentum smearing.

\subsection{HYP smearing}
As in our earlier study, we apply HYP smearing
\cite{Hasenfratz:2001hp} to the gauge links in the inserted
operator. This is a lattice technique used to smoothen the
gauge links, and such procedure is expected to bring the
renormalization factors closer to the corresponding
tree-level values.

This is particularly useful for the present status of our
work, because we still have not computed the operator
renormalization, while the renormalization factors for the local bilinear quark operators are
known: $Z_V=0.625(2)$,
$Z_A=0.7556(5)$, and $Z_T=0.7483(6)$ in the $\overline{\mathrm{MS}}$ scheme, at $\mu=2$ GeV, where we used the renormalization functions from Ref.~\cite{Alexandrou:2015sea}, which have also been employed for other nucleon structure quantities \cite{Abdel-Rehim:2015owa,Alexandrou:2013joa}.

In this way, in order to estimate how operator
renormalization could affect the present results, we apply 5
steps of HYP smearing, which, according to our previous
study, is close to saturating the effects of the smearing. In
fact, it is shown in Ref.~\cite{Ishikawa:2016znu} that
the linear divergences have their origin in the tadpole-type diagrams and
that HYP smearing of the Wilson line
has, practically, the effect of removing such linear divergences.

\subsection{Lattice setup}
\label{sec:setup}
\textbf{Ensemble.} We use a $32^3 \times 64$ ensemble from an ETMC (European
Twisted Mass Collaboration) production ensemble
\cite{Baron:2010bv}, with $N_f = 2+1+1$ flavors of
maximally twisted mass fermions. The bare coupling is $\beta
= 1.95$, corresponding to a lattice spacing of $a \approx
0.082$\,fm \cite{Carrasco:2014cwa}, while the twisted mass parameter is $a\mu = 0.0055$,
which gives a pion mass of $m_{PS}\approx 370$\,MeV.

\noindent\textbf{Statistics.} For the computation of matrix elements, we
first extend our previous calculation using Gaussian
smearing.  To that end, we now employ 1000 gauge
configurations, each with 15 point source forward
propagators and 2 stochastic propagators, resulting in total
30000 measurements, which is about 6 times more
than our previous paper \cite{Alexandrou:2015rja}.
For our exploratory computation using the momentum smearing method, we
use 50 gauge configurations for momentum $6\pi/L$ and
$8\pi/L$, and 100 gauge configurations for momentum
$10\pi/L$. In both cases, we use 3 sequential quark
propagators, one for each spatial direction, resulting in a
total of 150 measurements for $P_3 = 6\pi/L$ and $8\pi/L$,
and 300 measurements for $P_3=10\pi/L$.
As we will demonstrate below, even with this rather small statistics, we obtain good results thanks to the new momentum smearing.

\section{Numerical results}

In this section, we show our numerical results.
For purely illustrative purposes, we start with a simple free theory demonstration that indeed the expected quark distribution, the Dirac delta function at $x=1/3$, is approached when sufficiently increasing the nucleon momentum.
Then, we discuss the results for the matrix elements needed for the computation of quasi distributions.
In particular, we show the tremendous improvement from using the recently introduced momentum smearing.
Next, we move on to the quasi distributions and the physical parton distribution functions obtained from the matching procedure.
We also discuss the importance of HYP smearing and finally, we show results for the moments of the distributions determined in this work.

\subsection{Free quark distributions}

\begin{figure}
\centering
\includegraphics{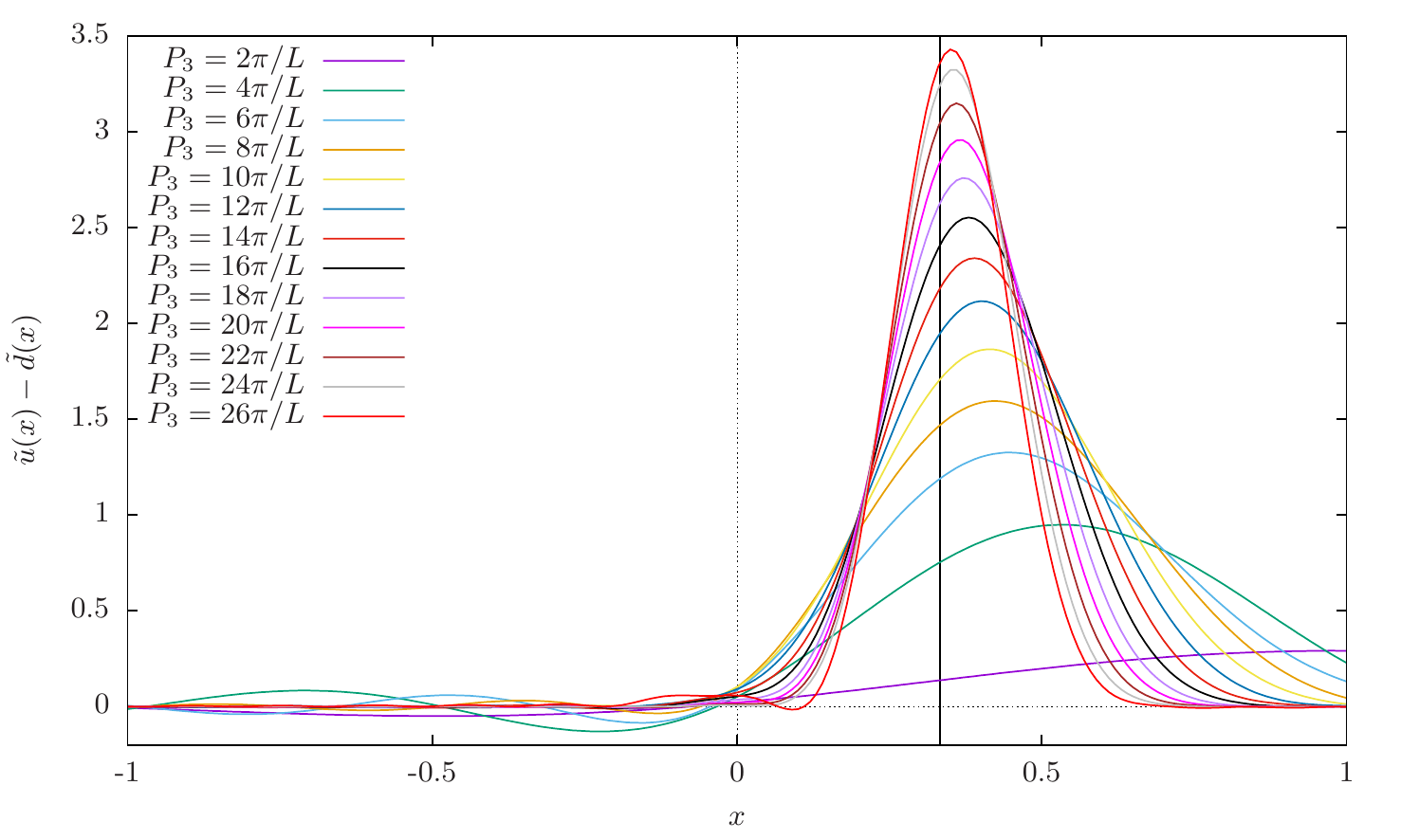}
\caption{\label{FIG_FREE_QUARKS} Free quark distributions for a lattice of size  $48^3\times96$. As expected, the distributions tend to a Dirac delta at $x=1/3$ as the nucleon momentum grows.}
\end{figure}

Free quark distributions are obtained from the definitions
(\ref{eq1}) or (\ref{eq2}) when one sets the gauge links
equal to 1. With unity gauge links, the limitation
connected to a poor signal-to-noise ratio disappears, as
well as the problem of renormalization, and, given this, the
free quark distributions are the perfect prototype to make explicit
the fundamental point  of the present approach: the purely spatial
correlations calculated in the lattice, the quasi-distributions, tend to
what we expect for the quark distributions as $P_3 \rightarrow \infty$.
If the
quasi-distribution approach is, thus, to be valid, the resulting
distributions should tend to a Dirac delta function,
centered in $x=1/3$, as $P_3$ increases.  We show in Fig.
\ref{FIG_FREE_QUARKS} our results for a lattice of size
$48^3\times96$, for a nucleon with momentum from
$P_3=2\pi/L$ to $P_3=26\pi/L$. The computations are done
with a source-sink separation of 12. The resulting
distributions are a very pedagogical illustration on the
foundations of the method for calculating quark
distributions discussed in this work. It shows clearly that
as the nucleon momentum increases, the probability of having
quarks carrying a negative fraction of the nucleon momentum,
or also carrying more momentum than the whole nucleon
itself, is suppressed to zero. Simultaneously, the shape of
the distributions changes to what is physically expected,
namely for large $P_3$ they start to become sharply
distributed around $1/3$.  The Dirac delta function should
be recovered in the combined continuum and infinite
source-sink separation limits. This is, however, beyond the
scope of the present work.

The results shown in Fig.~\ref{FIG_FREE_QUARKS} for the case
of free quarks are a fair indication that one should push the
calculation of the quasi distributions to values of the
nucleon momentum as high as computationally possible in the
lattice.

\subsection{Matrix elements}
\begin{figure}[t!]
\centering
\includegraphics{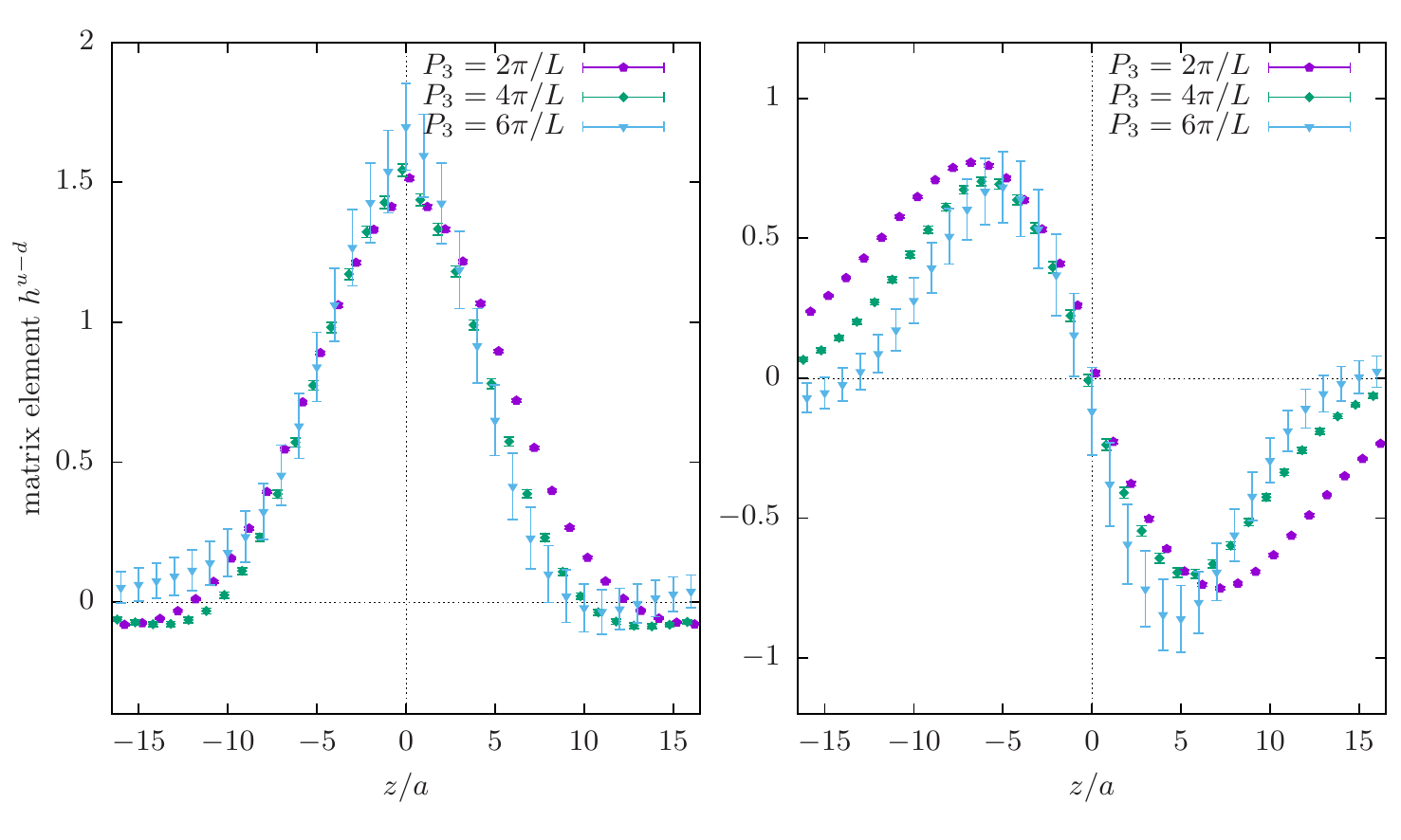}
\caption{\label{FIG_ME_GAUSS_UNPOL} Real (left) and imaginary (right)
parts of the matrix elements for the case of the vector operator.}
\end{figure}
\begin{figure}[t!]
\centering
\includegraphics{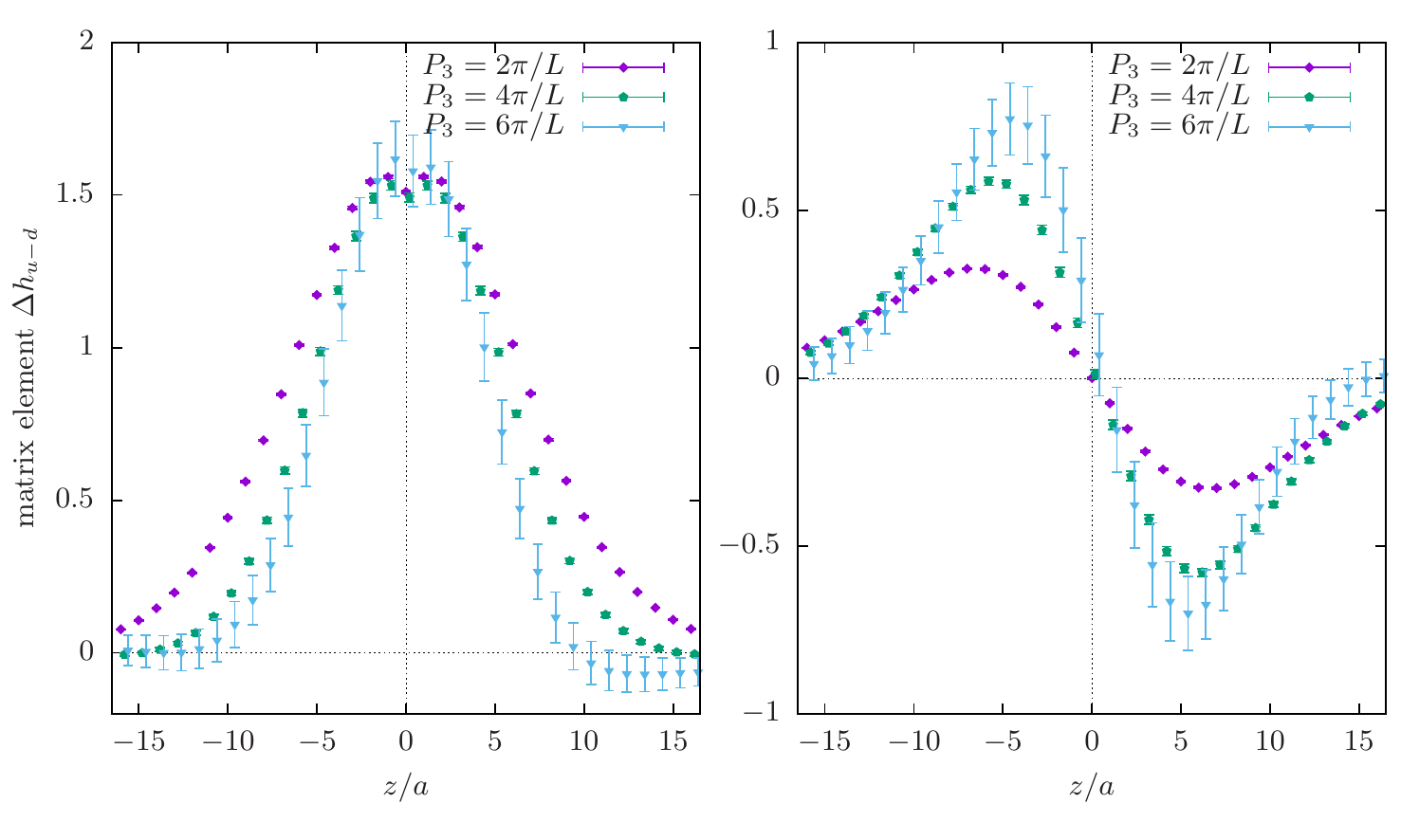}
\caption{\label{FIG_ME_GAUSS_HEL} Real (left) and imaginary (right)
parts of the matrix elements for the case of the
axial-vector operator.}
\end{figure}

In Figs.\,\ref{FIG_ME_GAUSS_UNPOL} and
\ref{FIG_ME_GAUSS_HEL}, we show the matrix elements for the
unpolarized $h^{u-d}(P_3,z)$ and the helicity $\Delta
h^{u - d}(P_3,z)$ cases, calculated for the 3
lowest lattice momenta, $P_3=2\pi/L, 4\pi/L,$ and $6\pi/L$, with Gaussian smearing and high statistics (see Sec.~\ref{sec:setup}).
For the configurations used here, these correspond, in
physical units, to $P_3 = 0.49$, $0.98$ and $1.47$\,GeV,
respectively. Compared to our previous result, the errors
are smaller by a factor of about 2.5, making them very
precise for the two lowest momenta, but still somewhat big
for $P_3=6\pi/L$, which imposes, as discussed in the
beginning of this section, a serious limitation.
We note that for lower momenta, $P_3=2\pi/L$ and $4\pi/L$, the matrix elements for the unpolarized case are not compatible with zero for the largest computed Wilson lines lengths $|z/a|=16$. This affects the normalization of the extracted PDFs, and indicates large corrections connected to the finite value of $P_3$, presumably too large to make already contact to the physical PDFs with these momenta, as clearly seen in the plots of the left side of Fig. 5 of Ref.~\cite{Alexandrou:2015rja}.
Thus, these momenta are too low to be taken into account in the extraction of PDFs, e.g.\ in the extrapolation to infinite momentum that will be part of the procedure when we aim at calculations to be compared to the phenomenological results (which we plan for the physical pion mass ensembles of ETMC and after addressing a suitable renormalization, see Sec.~\ref{sec:conclusions}). For this reason, a check of the normalization of the resulting PDFs, which is done in Sec.~\ref{sec:moments},  is very important.

To overcome the restriction on the values of the nucleon
momentum that can be used, we now employ the momentum
smearing on the quark fields, Eq.\,(\ref{momsmearing}), and
recalculate the matrix elements $h^{u-d}(P_3,z)$ and $\Delta
h^{u - d}(P_3,z)$. In practice, we use a Gaussian smearing
routine and include gauge links with a complex phase
$e^{ik\hat{j}}$. For now, we follow \cite{Bali:2016lva} and
choose $\zeta$ to be 0.45.

In addition, we also calculate the matrix
elements of the transversity operator, $\delta h^{u -
d}(P_3,z)$. The computation is done for momentum
$P_3=6\pi/L$, $8\pi/L$ and $10\pi/L$ for the case of
$h^{u-d}(P_3,z)$, and for $P_3=6\pi/L$ for the cases of
$\Delta h^{u - d}(P_3,z)$ and $\delta h^{u - d}(P_3,z)$.

\begin{figure}[t!]
\centering
\includegraphics{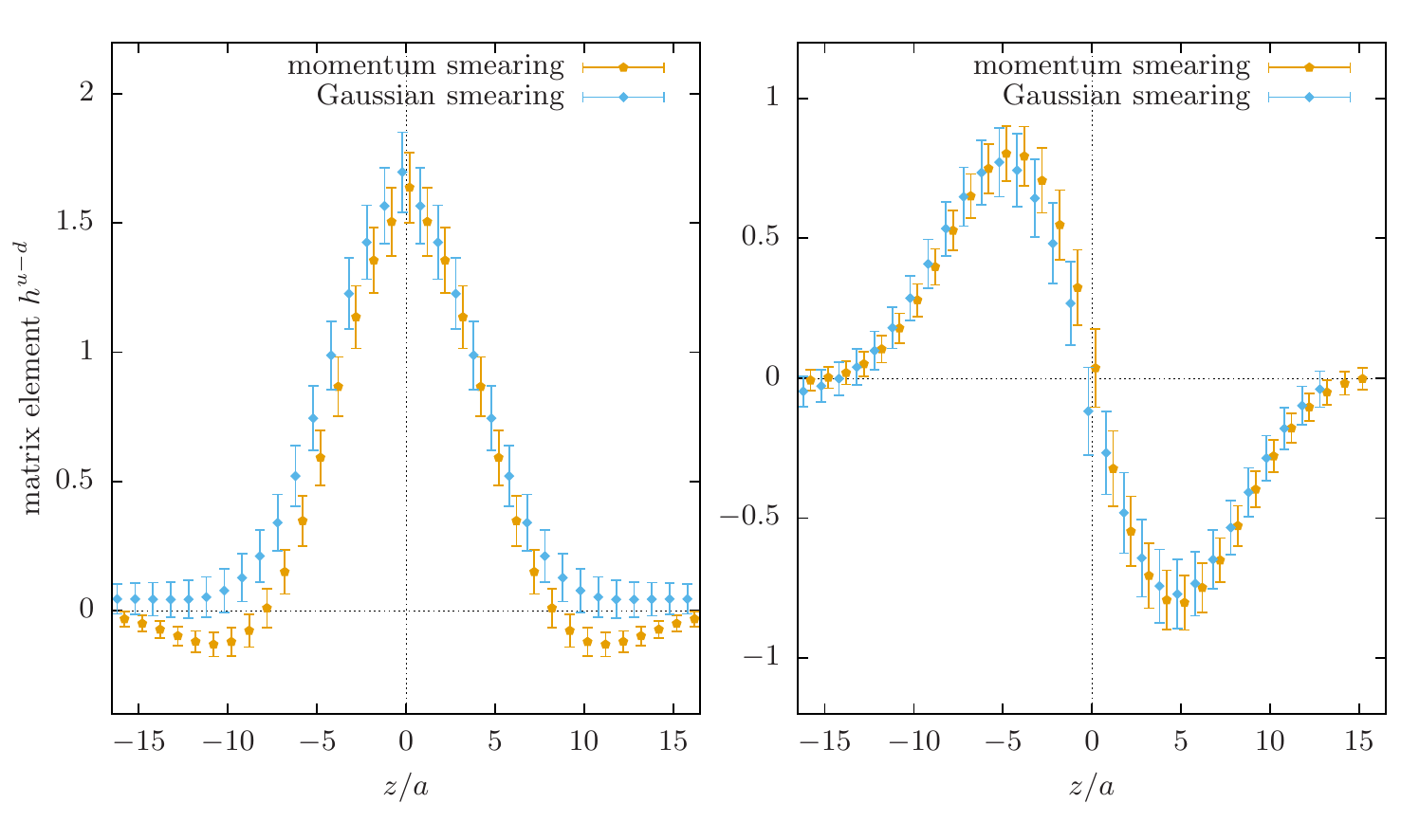}
\caption{\label{FIG_ME_COMPARE} Comparison of two different methods for the smearing
of the quark fields for the computation of the matrix element $h^{u - d}(P_3,z)$. One is the standard Gaussian smearing,
where 30000 measurements were used. The other is the new momentum smearing, where similar results to those
of Gaussian smearing are achieved using only 150 measurements, for the case
of momentum $P_3 =  6\pi/L$. In this plot, the real (imaginary) parts of the matrix elements were symmetrized (antisymmetrized).}
\end{figure}

We first compare the $h^{u - d}(P_3,z)$ matrix elements
using the two different approaches for the smearing of the
quark fields, for the case of $P_3=6\pi/L$ only. The result
is shown in Fig.\,\ref{FIG_ME_COMPARE} (in this case, we have symmetrized (antisymmetrized) the real (imaginary) parts of the matrix elements). The two different methods for
the smearing give compatible results for the matrix
elements, and it is remarkable that the number of
measurements necessary for the momentum smearing to match
the results from the Gaussian smearing is smaller by a
factor of 200.
The compatibility of the results from both smearings also suggests that the contamination by excited states is, in the analyzed matrix elements, very similar, i.e.\ it is very small at $P_3=6\pi/L$, as we have explicitly shown in our previous investigation \cite{Alexandrou:2015rja} by comparing two different source-sink separations. However, excited states contamination is expected to increase at higher $P_3$ \cite{Roberts:2012tp}. Hence, this systematic effect will be thoroughly investigated in future work aimed at extracting PDFs directly at the physical pion mass.

\begin{figure}[t!]
\centering
\includegraphics[width=0.8\textwidth]{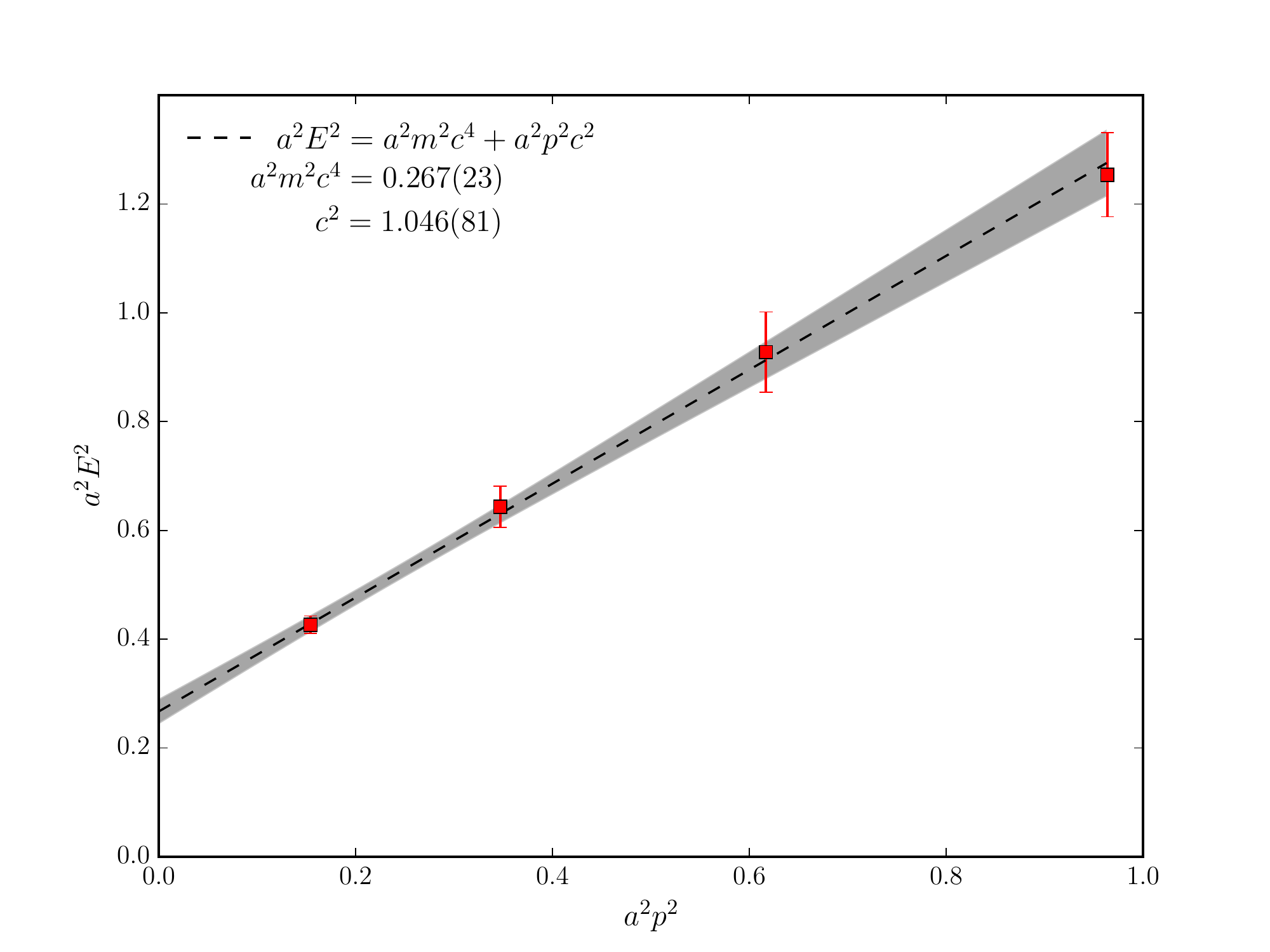}
\caption{\label{FIG_DISPERSION} Energy of the nucleon as a function of the momentum boost $p\equiv P_3$. We fit the relativistic relation between the energy and the momentum, $E^2=m^2c^4 + p^2c^2$ and find the speed of light $c^2=1.046(81)$ and the (squared) nucleon mass in lattice units $a^2m^2c^4=0.267(23)$.}
\end{figure}

We also show the continuum dispersion relation for the nucleon, using momenta $P_3=4\pi/L, 6\pi/L, 8\pi/L, 10\pi/L$ (all from momentum smearing).
We find that the relativistic relation between the energy and the momentum $p\equiv P_3$, $E^2=m^2c^4 + p^2c^2$ is satisfied, i.e. the fit of this relation to our lattice data describes the data very well and the fitted value of the speed of light is $c=1.023(40)$, i.e.\ it is compatible with the expected value of $c=1$ in our units. The nucleon mass in lattice units extracted from this fit, $amc^2=0.517(22)$, is also compatible with its direct extraction for a nucleon at rest, $amc^2=0.503(2)$ (with $c=1$) \cite{Alexandrou:2013joa}.

\begin{figure}[t!]
\centering
\includegraphics{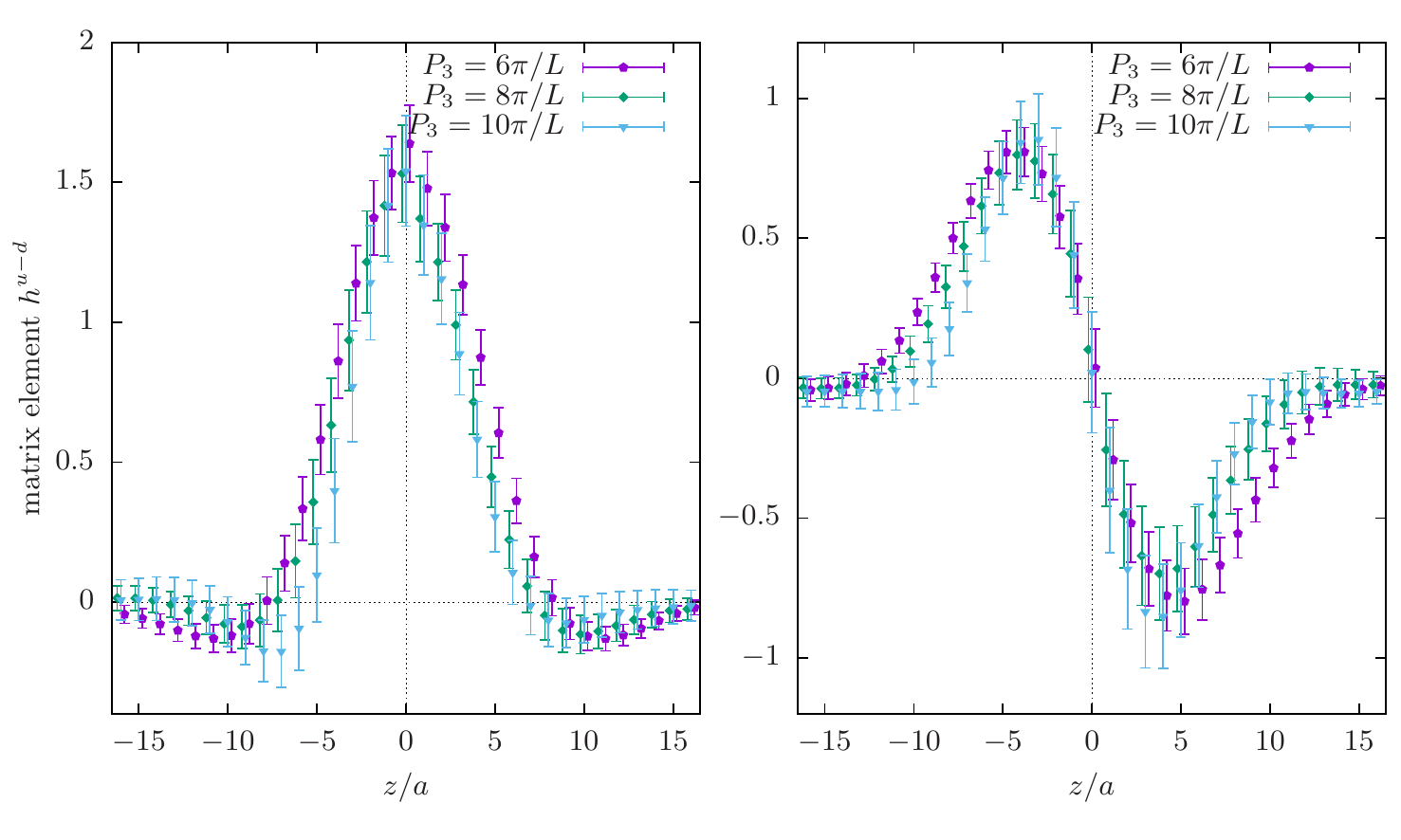}
\caption{\label{FIG_ME_MS} Matrix elements $h^{u-d}(P_3,z)$ (unpolarized operator) for momentum $P_3 =  6\pi/L, 8\pi/L,$ and $10\pi/L$, using momentum smearing.}
\end{figure}
\begin{figure}[t!]
\centering
\includegraphics{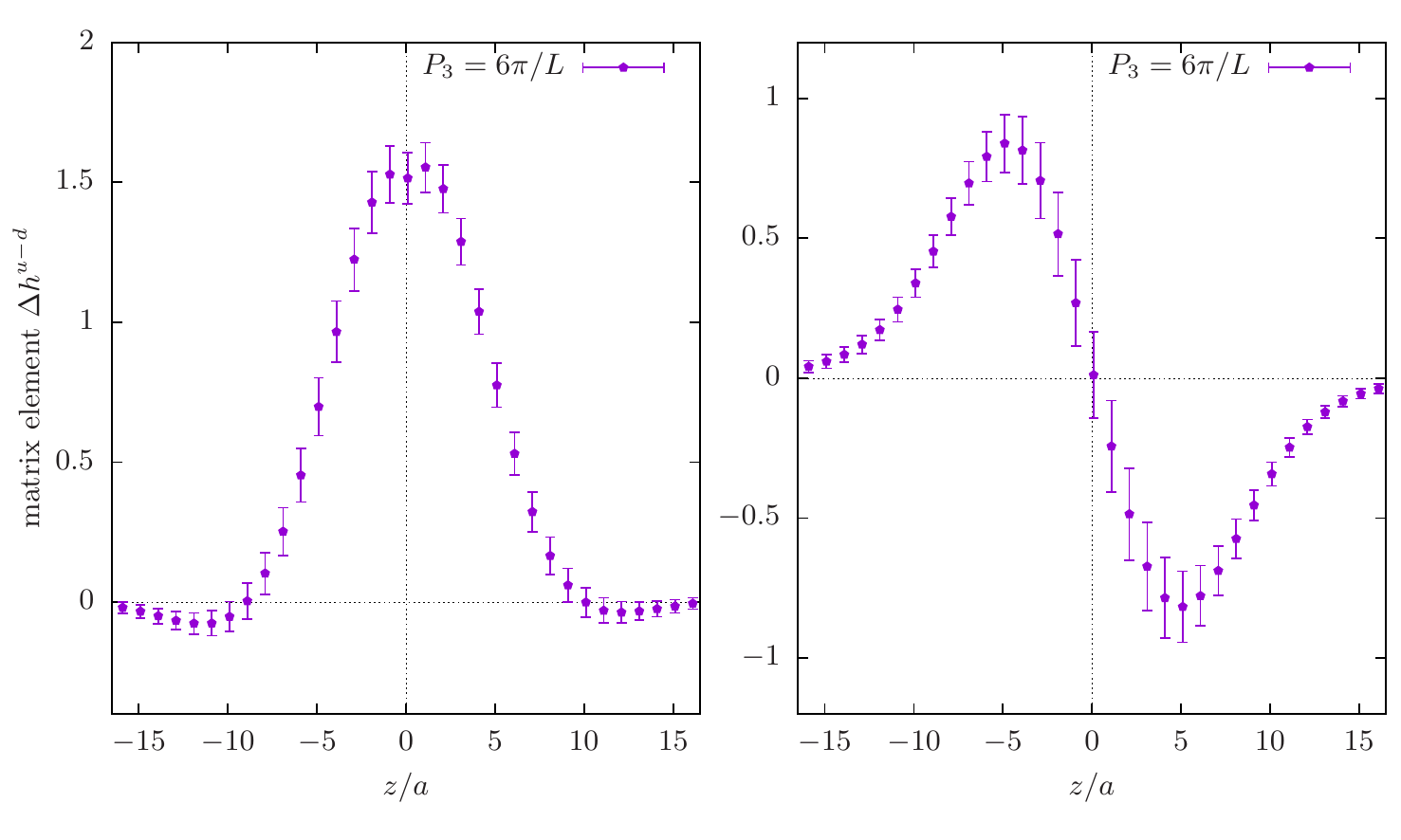}
\includegraphics{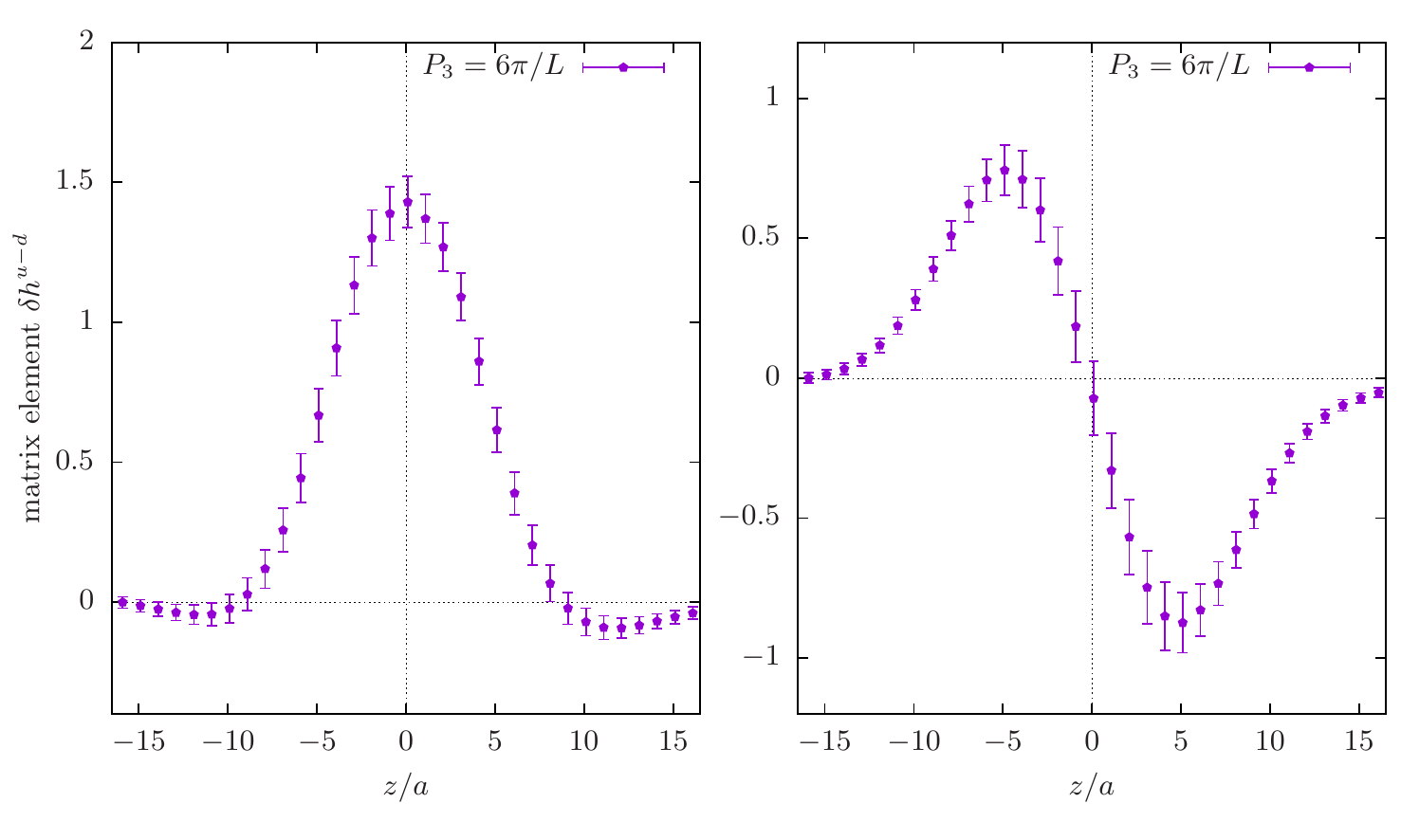}
\caption{\label{FIG_ME_MS_HEL} Matrix elements for the helicity and transversity operators for momentum $P_3 =  6\pi/L$, using momentum smearing
for the quark fields.}
\end{figure}

Having established the importance of momentum smearing for
the quark fields, we show in Fig.\,\ref{FIG_ME_MS} the matrix
elements $h^{u-d}(P_3,z)$ for $P_3 =  6\pi/L$, $8\pi/L$ and
$10\pi/L$. With this, we are now ready to calculate the
quasi distributions using Eq.\,(\ref{QuasiDistributions}),
and the corresponding quark distributions through
Eq.\,(\ref{invq}) after applying the TMCs. This is done
in the next subsection.

In Fig.\,\ref{FIG_ME_MS_HEL}, we
show results for the matrix elements of helicity $\Delta
h^{u - d}(P_3,z)$ and, for the first time in
our work, the transversity distribution $\delta h^{u -
d}(P_3,z)$ for $P_3 =  6\pi/L$
and the usage of momentum smearing.

\subsection{Quark distributions}
\label{SEC_QD}
\begin{figure}
\centering
\includegraphics{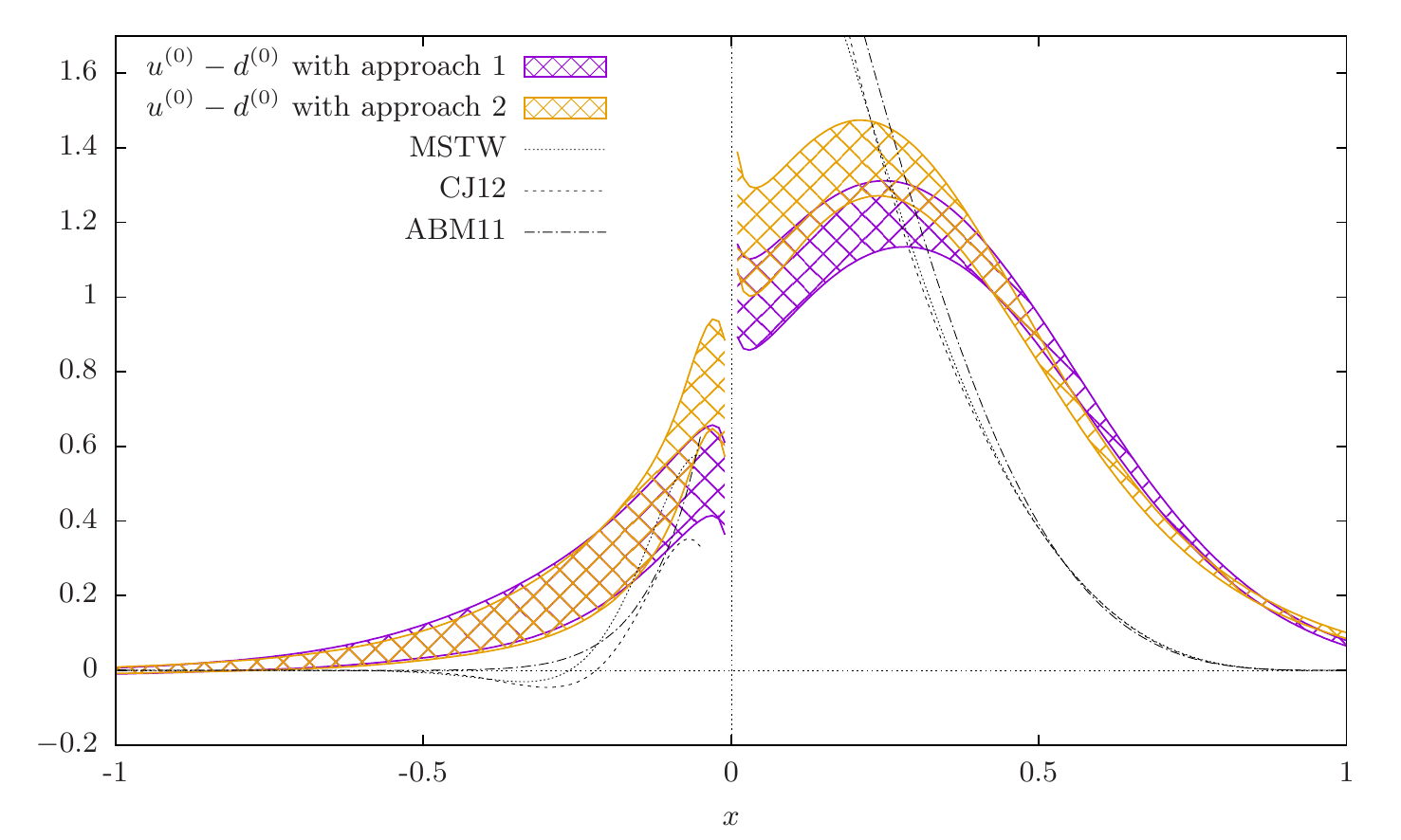}
\caption{\label{FIG_MS_COMPARE_MASS} Iso-vector quark distributions calculated with different prescriptions for the TMCs, for $P_3 =  6\pi/L$, using momentum smearing. Prescription (1) is
the one of Ref.~\cite{Alexandrou:2015rja}, while prescription (2) is the one of Ref.~\cite{Chen:2016utp}.}
\end{figure}

In this section, we compute the iso-vector quark
quasi distributions and the iso-vector quark distributions
in the nucleon, using the matrix elements calculated in the
previous subsection, together with Eqs.  (\ref{invq}) and
(\ref{QuasiDistributions}). We also apply the TMCs (for all employed Dirac structures, corresponding to the unpolarized, helicity and transversity cases) using for
them the prescription of \cite{Alexandrou:2015rja},
\begin{equation}
\tilde{q}(x,P_z) = \tilde{q}^{(0)}(\xi, P_z)/(1 + \nu \xi^2),
\label{TMC}
\end{equation}
where $\xi = 2x/(1+\sqrt{1+4\nu x^2})$ is the Nachtmann
variable, $\nu = M_N^2/4(P_3)^2$ ($M_N$ -- nucleon mass), and the superscript $(0)$ means that the TMCs have
been taken into account.  In Ref.~\cite{Chen:2016utp}, an
improved prescription for TMCs, which avoids the problem of
non-preservation of the norm of the distributions, was
presented. However, as shown in Fig.
\ref{FIG_MS_COMPARE_MASS}, for the values of the nucleon
momentum that are the main interest of this work ($P_3 \geq
6\pi/L$) there is, in practice, no difference between the
two prescriptions, mostly in the intermediate and large $x$
regions, which are the regions where the quasi distributions
approach is applicable. For $P_3=8\pi/L$, the difference
between the two prescriptions is essentially non existing
for the whole $x$ region. We also do not apply the
corrections in $\Lambda_{QCD}/P_3$, because of the large
values of $P_3$ used in this work to compute the
quasi distributions.

For the UV cutoff, we use $\Lambda=1/a \approx 2.5$ GeV,
this being also our choice for the renormalization scale
$\mu$ where the distributions are defined.  For the bare
coupling, we use $\alpha_s = 6/(4\pi\beta)$, which, for
the lattice setup employed here, corresponds to $\alpha_s
\approx 0.245$. Of course, in the future, a full
renormalization program will have to be carried out,
implying the independence of the final results on $\Lambda$.
In this same line, we also notice that the integrals in Eq.\,(\ref{invq}) are cutoff dependent through their dependence
on $x_c \sim \Lambda/P_3$.  This cut in $x$, $x_c$, marks
the region in $x$ where $\tilde{q}(x > x_c, \Lambda, P_3) =
0$. Of course, $x_c \geq 1$ by construction, since the nucleon boost can not be larger than the UV cutoff.

\subsubsection{Gaussian smearing}

\begin{figure}
\centering
\includegraphics{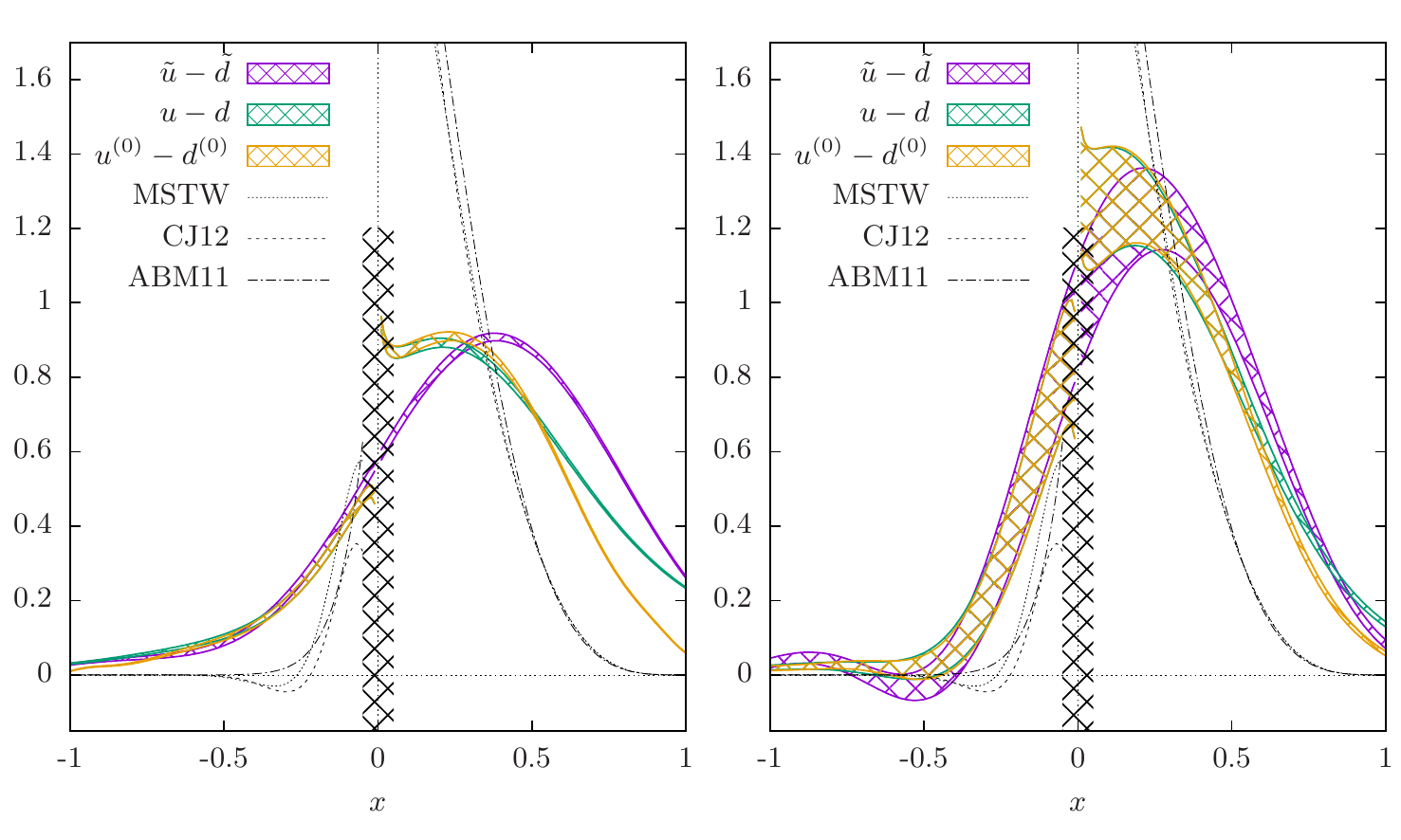}
\includegraphics{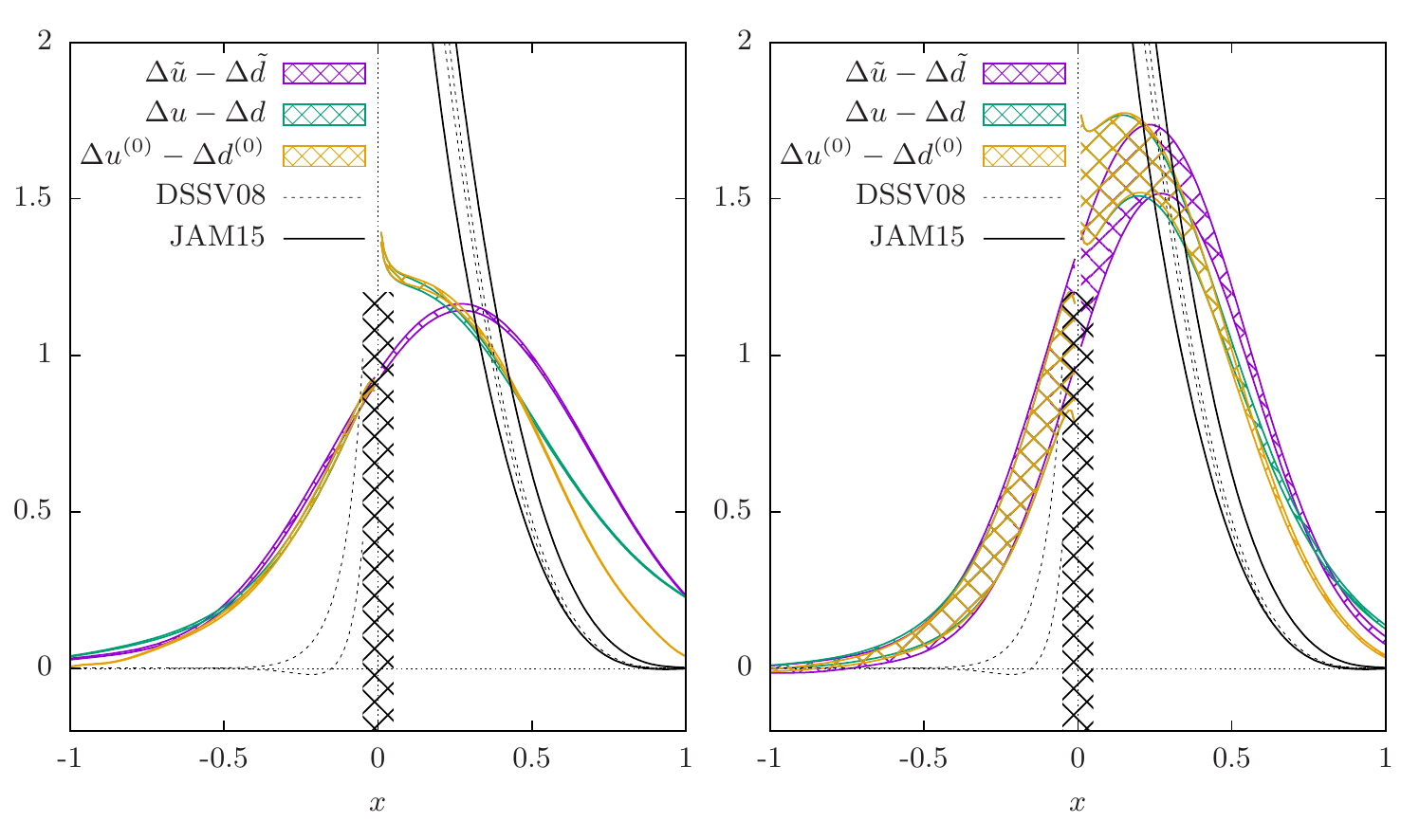}
\caption{\label{PDF_GAUSS} The quark quasi distributions ($\tilde q$), the quark distributions ($q$), and the quark distributions after TMCs are applied ($q^{(0)}$), for the unpolarized and
helicity iso-vector combinations, and for $P_3 =  4\pi/L$ (left) and $6\pi/L$ (right). The standard Gaussian smearing for the quark fields is used.}
\end{figure}

With these chosen parameters, we show in Fig.
\ref{PDF_GAUSS}  the results for the quasi distributions,
$\tilde q(x)$, the distributions before mass corrections,
$q(x)$, and the distributions after mass corrections,
$q^{(0)}$, for $P_3 = 4\pi/L$, and $6\pi/L$, for both the
unpolarized and helicity iso-vector quark distributions, with Gaussian smearing.
The phenomenological parameterizations for the experimental
data are taken from MSTW \cite{Martin:2009iq}, CJ12
\cite{Owens:2012bv}, and ABM11 \cite{Alekhin:2012ig}, for
the unpolarized case; for the helicity distributions, the
parameterizations are taken from DSSV08
\cite{deFlorian:2009vb} and JAM15 \cite{Sato:2016tuz}. In
both cases, we observe a tendency of the calculated
distributions to move towards the parameterizations as the
nucleon momentum increases, as expected.  Also, it is clear
from the plots that both the matching and the TMCs become
less relevant as the momentum increases, a key observation
for the analysis of the data for momentum $8\pi/L$ and
$10\pi/L$, as will be done in this section.
Our results are compatible with those of Ref.
\cite{Chen:2016utp}. In particular, we also predict\footnote{The work of Schreiber, Signal, and Thomas \cite{Schreiber:1991tc} was the first one to predict, in the context of a bag model calculation, that $\Delta\overline u(x) - \Delta\overline d(x) > 0$.} that
$\Delta\overline u(x) - \Delta\overline d(x) > 0$, and such
asymmetry also seems to be experimentally observed by the
STAR \cite{Adamczyk:2014xyw} and PHENIX collaborations \cite{Adare:2015gsd}.

Having confirmed our former results for the unpolarized
distributions using 6 times more measurements than before,
and also having calculated the helicity distributions for
the first time in our framework, we are ready to test the
feasibility of extending such calculations for higher values
of the nucleon momentum.  Our results indicate that the use
of larger values for the momentum improves the agreement
between the lattice calculation and the parameterizations,
and now this can be reliably tested.

\subsubsection{Momentum smearing}

\begin{figure}
\centering
\includegraphics[scale=0.8]{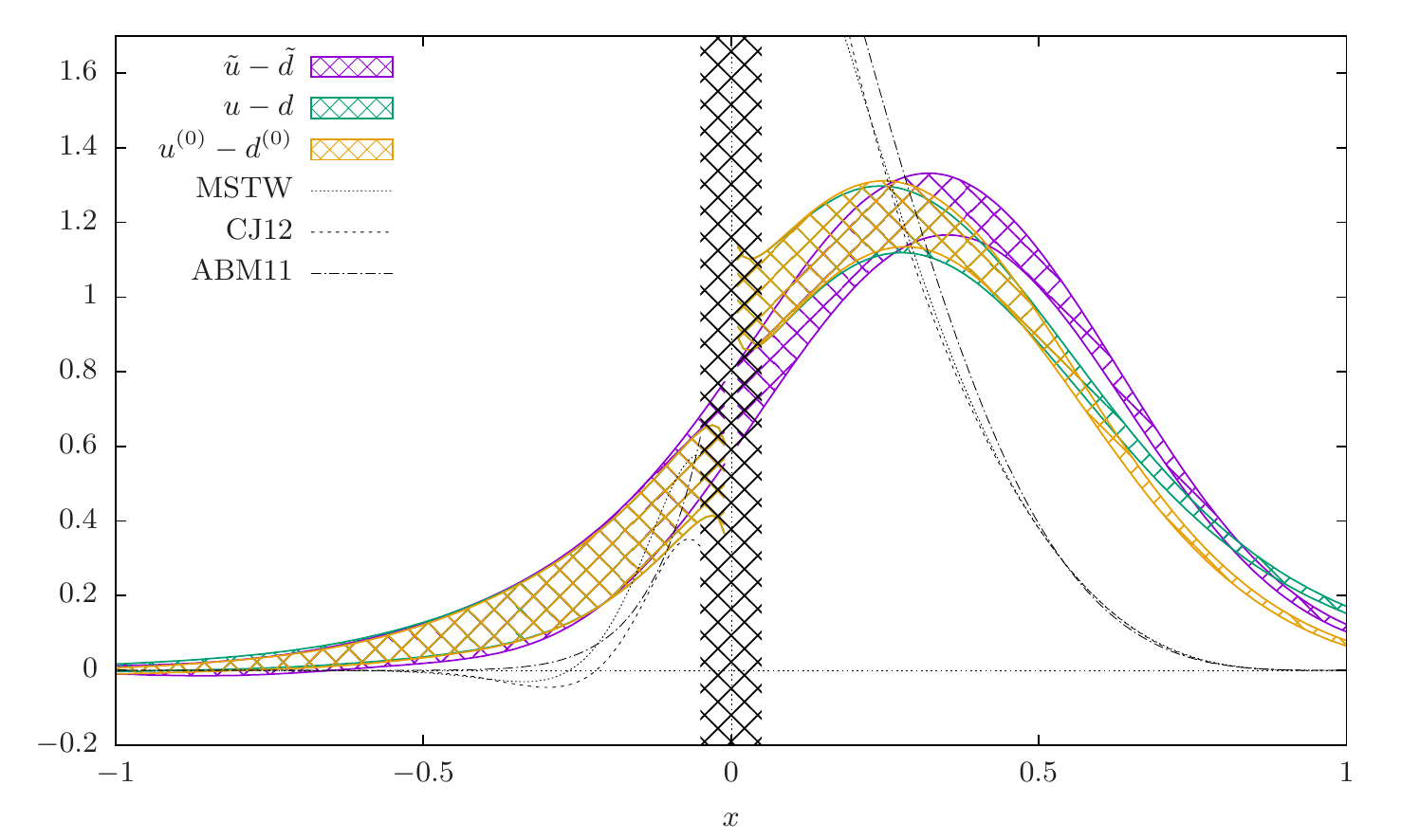}
\includegraphics[scale=0.8]{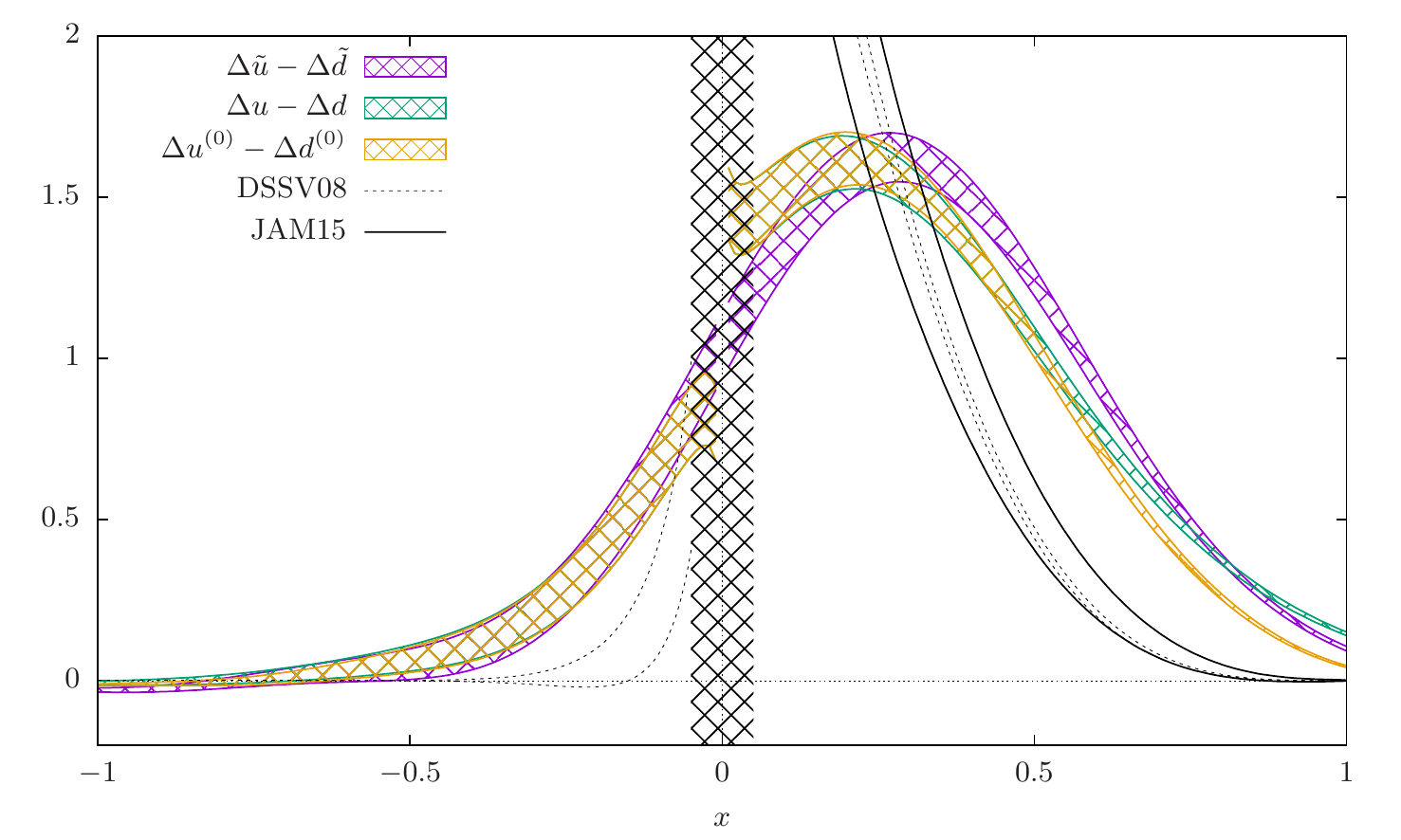}
\includegraphics[scale=0.8]{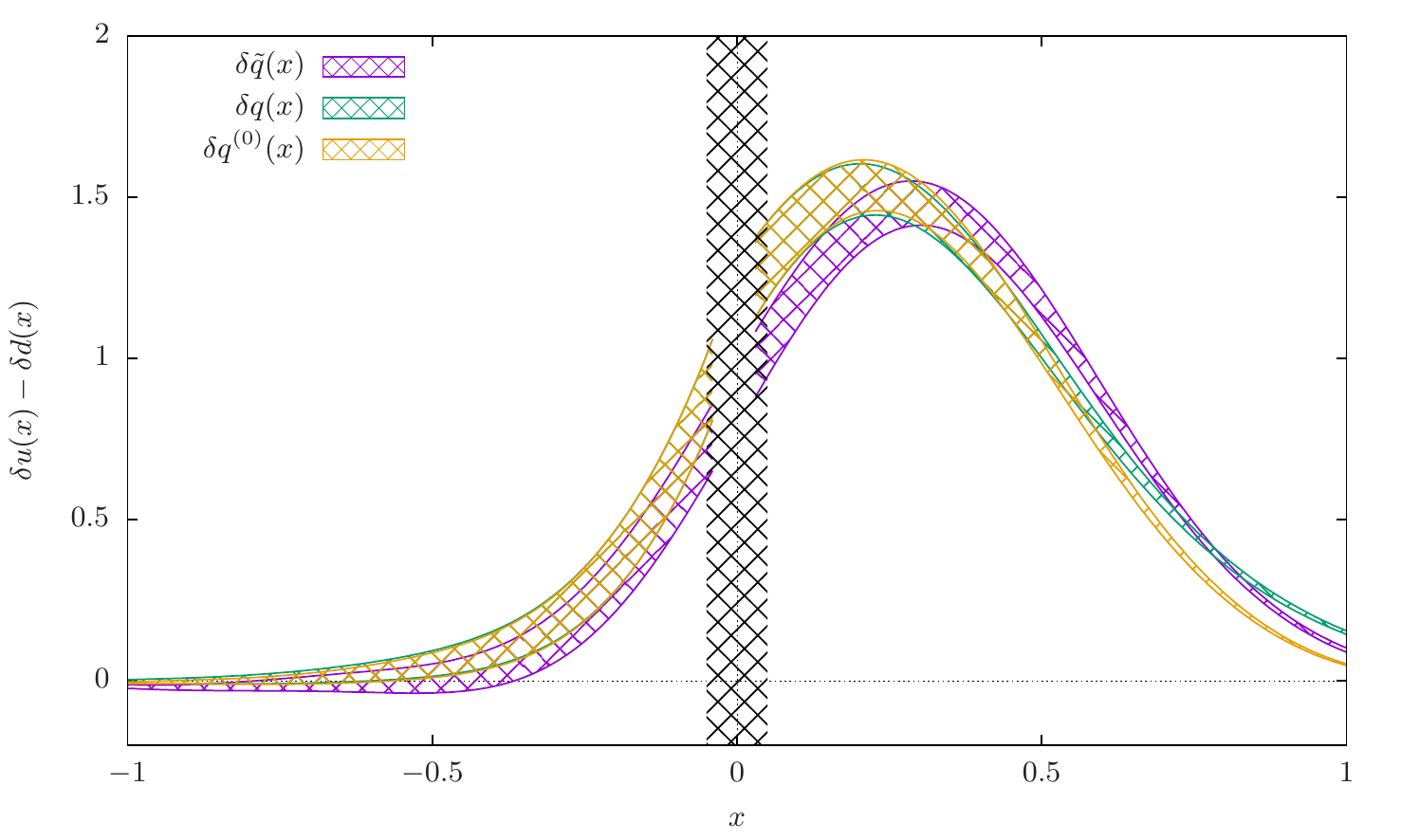}
\caption{\label{PDF_MS} The quark quasi distributions ($\tilde q$), the quark distributions ($q$), and the quark distributions after TMCs are applied ($q^{(0)}$), for the unpolarized,
the helicity, and the transversity iso-vector combinations, for $P_3 =  6\pi/L$. The new momentum smearing for the quark fields is used.}
\end{figure}

For the case of momentum smearing, we first recalculate the
quasi distributions and the distributions for
$P_3=6\pi/L$, in order to show that the new results are
compatible with those calculated with Gaussian smearing in Fig.
\ref{PDF_GAUSS}. The resulting curves, after applying the
matching and the TMCs, are shown in Fig.~\ref{PDF_MS}, where
we also present our results for the transversity distributions. We
do not show the parameterizations for the transversity distributions,
because they are largely unconstrained (see, for instance, Fig.~3 of
Ref.~\cite{Kang:2015msa}).
Future JLab data \cite{JLAB_TRAN}, however, will help to constrain the shape
of $\delta u(x)$ and $\delta d(x)$, and precise lattice data can
be a valuable guidance in that quest.

\begin{figure}
\centering
\includegraphics{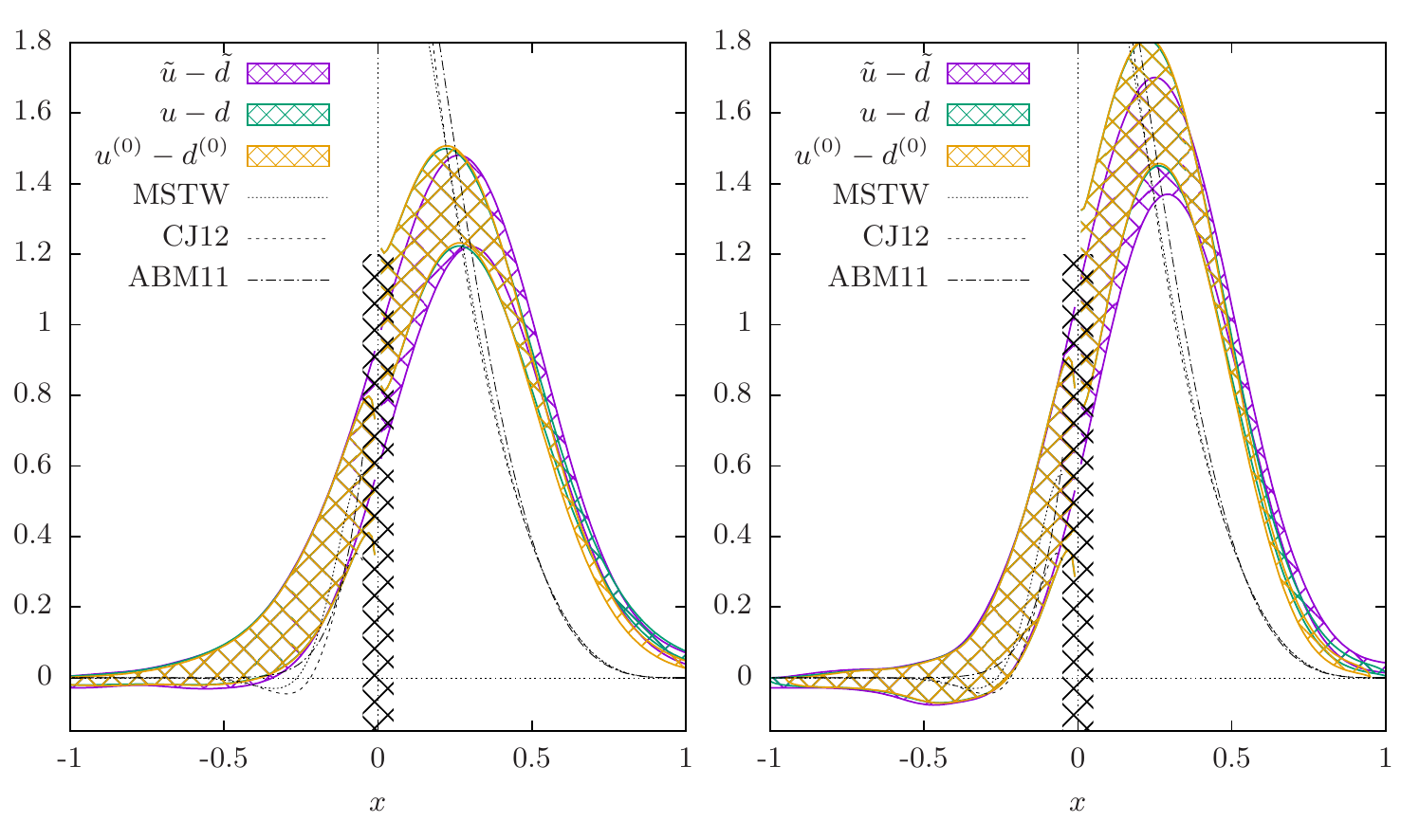}
\caption{\label{PDF_MS_UNPOL_M4M5} The quark quasi distributions ($\tilde q$), the quark distributions ($q$), and the quark distributions after TMCs are applied ($q^{(0)}$), for the unpolarized iso-vector distributions, and for $P_3 =  8\pi/L$ (left) and $10\pi/L$ (right). Momentum smearing for the quark fields is used.}
\end{figure}

\begin{figure}
\centering
\includegraphics{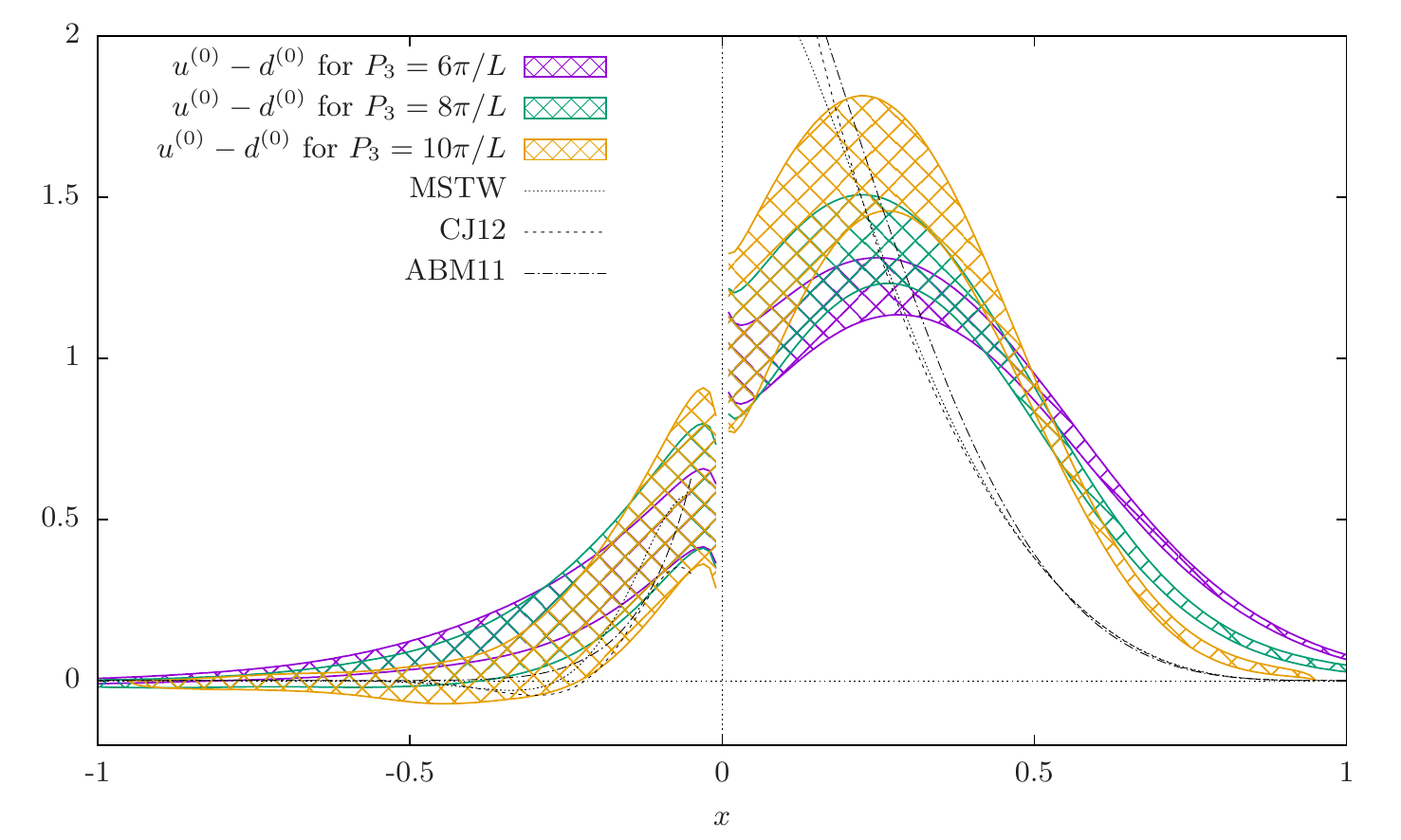}
\caption{\label{FIG_PDF_MS_COMPARE_MOM_q0} The unpolarized iso-vector quark distributions for momenta $P_3=6\pi/L, 8\pi/L$ and $10\pi/L$.}
\end{figure}

We are now ready to show our outcome for $P_3 = 8\pi/L$ and
$10\pi/L$, and that is done in Fig.~\ref{PDF_MS_UNPOL_M4M5}.
It is clear from this figure that the matching, and the
TMCs, provide almost no corrections to the
quasi distributions as the nucleon momentum grows, meaning
that for $P_3=10\pi/L$, the quasi distributions are,
essentially, the distributions themselves. This is true
mainly in the intermediate and  large $x$ regions, the
regions where the computation of the quasi distributions in
the lattice is valid: we know, from the uncertainty
principle relations, that $x_{min}\sim \Lambda_{QCD}/P_3$ is
the smallest value for $x$ where quasi distributions can be
reliably computed. Moreover, from the data, we also see that
the change in the resulting distributions with the nucleon
momentum is becoming smaller as $P_3$ increases. We
exemplify this by showing in Fig.~\ref{FIG_PDF_MS_COMPARE_MOM_q0}
the resulting quark distributions for $P_3=6\pi/L, 8\pi/L$ and $10\pi/L$, in one plot.
Figures \ref{PDF_MS_UNPOL_M4M5} and \ref{FIG_PDF_MS_COMPARE_MOM_q0}
encapsulate the important result that we can now compute the
quasi-distributions on the lattice for a proton with momentum large enough such
that the matching and the mass corrections are significantly smaller than previously
possible. Moreover, the resulting distributions have the correct support in $x$, and are consistent with zero
for $x\geq 1$ for the largest proton momentum studied here.

\subsection{HYP smearing and renormalization}
\label{sec:HYP}
The natural question to ask at this point is about the influence of renormalization on the results.
As we have emphasized above, our matrix elements have not been properly renormalized yet.
However, HYP smearing of the Wilson line involved in the computation of the three-point function is expected to bring the values of the renormalization constants of the resulting matrix elements closer to their tree-level values \cite{Capitani:2006ni}.
We demonstrate in this subsection that the effect brought in by HYP smearing leads to an improvement by bringing the extracted PDFs closer to the phenomenological curves.

\begin{figure}[t!]
\centering
\includegraphics{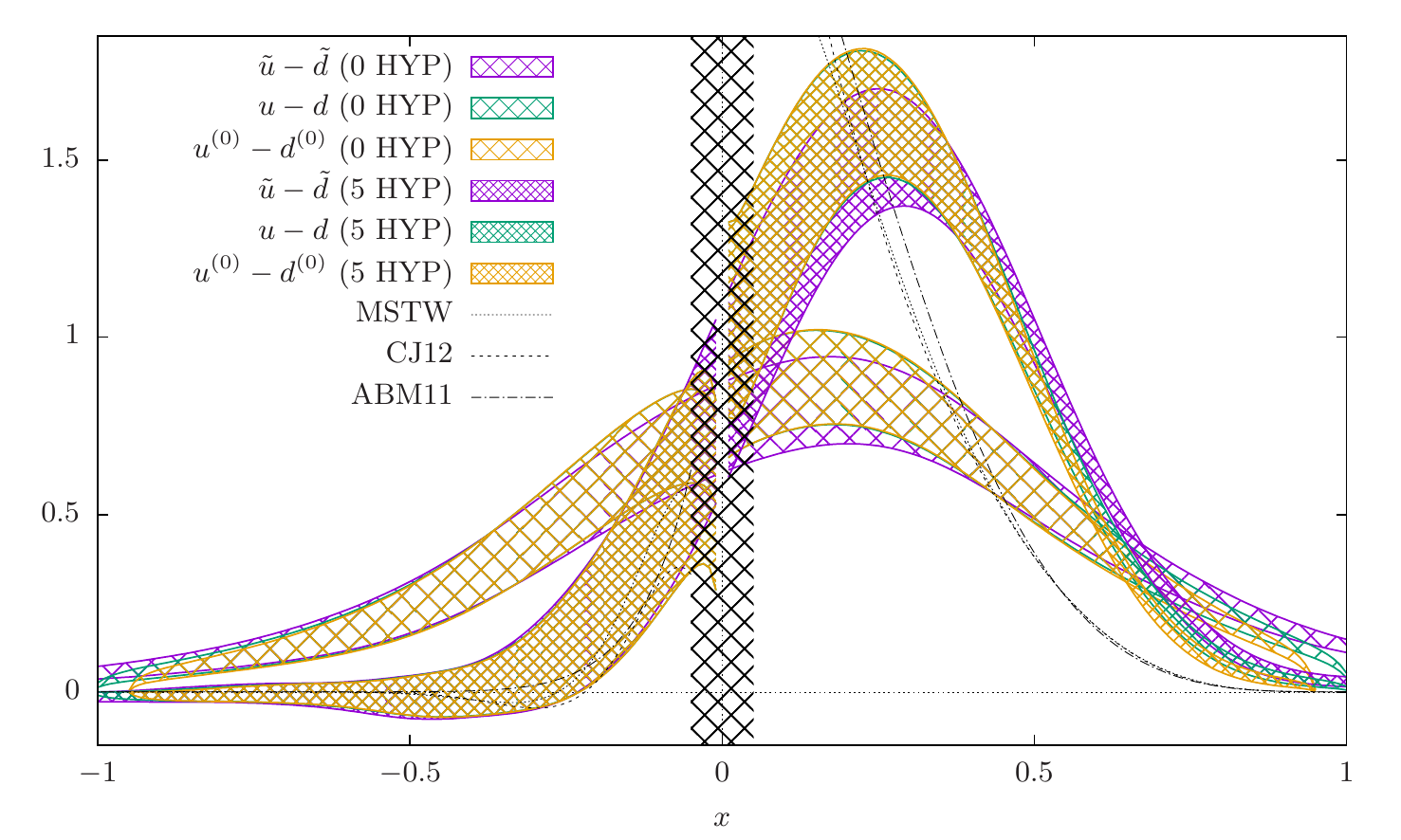}
\caption{\label{FIG_PDF_ms_unpol_HYP0} The unpolarized iso-vector quark distributions for momenta $P_3=10\pi/L$, with no HYP smearing in the gauge links (thinly filled curves) and with 5 steps of HYP smearing applied (thickly filled curves). Momentum
smearing for the quark fields was used. The comparison of 0 and 5 HYP smearing steps shows the importance of renormalization in producing an asymmetry between the $x$ and $-x$ regions, i.e.\ between quarks and antiquarks.}
\end{figure}

It is relevant to notice that because the matrix elements
are even under the interchange of the positive and the
negative $z$ regions and after taking the Hermitian
conjugate, as written in Eq.~(\ref{Asymmetry}), the
imaginary part of the matrix elements is odd under the
operation $z\rightarrow -z$.  When performing the Fourier
transform, Eq.\,(\ref{QuasiDistributions}), an asymmetry
between the positive and the negative $x$ regions appears
exactly because the imaginary part is an odd function. The
size of the resulting asymmetry is dependent on how large
the imaginary part of the matrix elements is. Moreover, as
shown in \cite{Alexandrou:2015rja}, using smaller values for the
nucleon momentum, a sizable imaginary part
only appears after HYP smearing is applied, suggesting that
renormalization of the inserted operator is fundamental to
produce a difference between the positive and the negative
$x$ regions. This
conclusion is corroborated by Fig.~\ref{FIG_PDF_ms_unpol_HYP0},
where we compare the unpolarized iso-vector quark distributions for
the case where no HYP smearing in the gluon fields is applied and when 5 steps of HYP smearing are used.
Without HYP smearing, there is only a very small asymmetry between the
positive and the negative $x$ regions, even if the momentum
used for the computation is $P_3=10\pi/L$.
Finally, because quarks in the negative $x$
region correspond to antiquarks in the positive $x$ region,
$\overline{q}(x) = -q(-x)$ for the unpolarized case,
$\Delta\overline{q}(x) = \Delta q(-x)$ for the helicity
case, and $\delta\overline{q}(x) = -\delta q(-x)$ for the
transversity case, the asymmetry between quarks and antiquarks in the
nucleon, using lattice QCD, is a direct consequence of the relations
(\ref{Asymmetry}).

As a final note on this subject, we stress that the effect of HYP smearing on the distributions
suggests that the properly renormalized results 
will be closer to the physical PDFs,
as compared to the unrenormalized PDFs.
However, it is important to point out that the missing renormalization is not the only reason why the phenomenological curves are not reproduced at this stage.
In addition to renormalization, there are other lattice effects that need to be controlled, such as cut-off effects and pion mass effects.
We provide some evidence for the importance of the latter in the next subsection.

\subsection{Moments of quark distributions}
\label{sec:moments}
We have also computed the moments of the quark distributions, i.e.\ quantities that were accessible to earlier lattice investigations, where they are extracted from matrix elements of local currents, e.g.\ from the forward matrix element of the one-derivative vector current for the first moment of the unpolarized PDF (see Refs.~\cite{Constantinou:2014tga, Constantinou:2015agp,
Alexandrou:2015yqa, Alexandrou:2015xts, Syritsyn:2014saa} for recent reviews).

\begin{table}[t!]
\begin{center}
\begin{tabular}{|c|c|c|c|c|c|c|c|}
\hline
\multirow{2}*{PDF type} & \multirow{2}*{$P_3$} & \multirow{2}*{Smear} & \multicolumn{3}{|c|}{Normalization} & \multirow{2}*{$\langle x\rangle_q$} & \multirow{2}*{$\langle x^2\rangle_q$} \\
\cline{4-6}
& & & antiquarks & quarks & total & & \\
\hline
\multirow{4}*{unpolarized} & \multirow{2}*{3} & Gauss & 0.187(55) & 0.752(56) & 0.94(11) & 0.219(28) & 0.134(12)\\
\cline{3-8}
 & & mom & 0.145(55) & 0.750(53) & 0.90(11) & 0.240(32) & 0.147(15)\\
\cline{2-8}
 & 4 & mom & 0.130(77) & 0.743(78) & 0.87(15) & 0.224(43) & 0.116(20)\\
\cline{2-8}
 & 5 & mom & 0.100(88) & 0.798(98) & 0.90(10) & 0.234(46) & 0.100(19)\\
\hline
\multirow{2}*{helicity} & \multirow{2}*{3} & Gauss & 0.253(62) & 0.920(58) & 1.17(12) & 0.249(29) & 0.154(12)\\
\cline{3-8}
 & & mom & 0.184(47) & 0.931(44) & 1.11(9) & 0.281(26) & 0.154(11)\\
\hline
\multirow{2}*{transversity} & \multirow{2}*{3} & Gauss & 0.175(99) & 0.923(95) & 1.10(19) & 0.309(67) & 0.163(35)\\
\cline{3-8}
 & & mom & 0.169(47) & 0.878(44) & 1.05(9) & 0.276(26) & 0.152(11)\\
\hline
\end{tabular}
\caption{\label{tab:moments} Moments of the computed PDFs. We show the PDF type, the value of $P_3$ (in units of $2\pi/L$), the smearing type used (Gauss=Gaussian, mom=momentum), the normalization (zeroth moment) decomposed to give contributions from quarks and antiquarks, the first moment and the second moment.
The normalization of helicity and transversity PDFs can be compared to the values of, respectively, $g_A^{u-d}=1.17(2)$ and $g_T^{u-d}=1.08(3)$ obtained by the ETMC for the same ensemble of gauge field configurations in Refs.~\cite{Alexandrou:2013joa,Abdel-Rehim:2015owa}.
The first moment values obtained in the same references are: $\langle x \rangle_q=0.233(9)$ (unpolarized), $\langle x \rangle_{\Delta q}=0.298(8)$ (helicity) and $\langle x \rangle_{\delta q}=0.316(12)$ (transversity).
}
\end{center}
\end{table}

We define the $n$-th moment in the following way.
For the unpolarized case,
\begin{equation}
\langle x^n \rangle_q = \int_{-1}^1 dx\,x^n q(x),
\end{equation}
where $q(x)$ is the unpolarized iso-vector quark distribution considered in this work. Similar expressions
hold for the helicity and transversity distributions. The 0-th moment should be equal to 1 for the unpolarized case. For the
helicity case, it should equal the iso-vector axial charge $g_A^{u-d}$, and for the transversity case the
iso-vector tensor charge $g_T^{u-d}$.
We also consider the decomposition of the moments into quark and antiquark parts, i.e.\ the splitting into the integral over negative (antiquarks) and positive (quarks) values of $x$ in the above formula.
For the antiquarks, it corresponds to the following integrals over positive $x$,
\begin{equation}
\langle x^n \rangle_{\bar q} = \int_0^1 dx\,x^n \left(\bar{d}(x) - \bar{u}(x)\right),
\end{equation}
\begin{equation}
\langle x^n \rangle_{\Delta \bar q} = \int_0^1 dx\,x^n \left(\Delta \bar{u}(x) - \Delta\bar{d}(x)\right),
\end{equation}
\begin{equation}
\langle x^n \rangle_{\delta \bar q} = \int_0^1 dx\,x^n \left(\delta \bar{d}(x) - \delta\bar{u}(x)\right),
\end{equation}
for the unpolarized, helicity and transversity cases, respectively, and where we have used the crossing relations introduced previously in Sec.~\ref{sec:HYP} \cite{Chang:2014jba}.

Our results, obtained from PDFs computed with 5 HYP smearing steps, are shown in Tab.~\ref{tab:moments}.
With this computation, we further test the hypothesis that HYP smearing can serve as a crude substitute for renormalization (although not replace it).
If such extracted moments agree with the renormalized moments extracted directly, as it will turn out to be, such a hypothesis is further validated.

We find that the normalization condition (total $\langle x^0\rangle=1$) is always satisfied (within errors) for the unpolarized case. It is also worth noticing that the integral of the
antiquark asymmetry is also compatible with the experimental results (see \cite{Chang:2014jba} for a compilation of the experimental and phenomenological
results for the light quark sea asymmetry). For example, the New Muon Collaboration \cite{Arneodo:1994sh} finds that $\int_0^1 x^n \left(\bar{d}(x) - \bar{u}(x)\right) = 0.148(39)$.
However, obviously, one can not have a final conclusion based on these numbers, since the PDFs obtained in this work are not rigorously renormalized and the pion mass is non-physical.
For the normalization of the helicity and transversity PDFs, we find agreement with ETMC results for the same ensemble of gauge field configurations \cite{Alexandrou:2013joa} (with an update in Ref.~\cite{Abdel-Rehim:2015owa}), $g_A^{u-d}=1.17(2)$ and $g_T^{u-d}=1.08(3)$ (at source-sink separation of $16a$).

As for the first moment, the results from ETMC's direct extraction are: $\langle x \rangle_q=0.233(9)$ (unpolarized), $\langle x \rangle_{\Delta q}=0.298(8)$ (helicity) and $\langle x \rangle_{\delta q}=0.316(12)$ (transversity).
These values are in a rather good agreement with the ones extracted here, even if they were obtained using a completely different approach.
Of course, the point of this comparison is not the expectation that the values will precisely coincide, but rather the qualitative feature that working at this non-physical pion mass, approx. 370 MeV, the quark momentum fraction is significantly above the phenomenological value, around 0.16/0.20 for the unpolarized/helicity case.
Conversely, ETMC computations at the physical pion mass lead to the values: $\langle x \rangle_q=0.208(24)$, $\langle x \rangle_{\Delta q}=0.229(30)$ and $\langle x \rangle_{\delta q}=0.306(29)$ \cite{Abdel-Rehim:2015owa}, i.e.\ much closer to the phenomenological values (for cases where they are precisely known).
This strengthens our expectation that working at the physical pion mass, the PDFs determined from the quasi-PDF approach should also be much closer to the phenomenological curves, i.e.\ it hints that the pion mass (in addition to the missing renormalization) is to a large extent responsible for the current shape of the extracted PDFs.
Nevertheless, other systematic effects, in particular cut-off effects may also play their role.

We also show values for the second moment, $\langle x^2 \rangle_q$. The point of this calculation is to demonstrate that higher moments are obtained with a similar relative precision as the first moments, since their uncertainty depends only on the relative error in the extracted PDFs. This is in opposition to the situation in direct moments extraction in the traditional lattice approach, where higher moments become notoriously difficult beyond the second or third moment and even those can be obtained with a much worse precision, due to a decreasing signal-to-noise ratio and complicated mixing patterns under renormalization.

\section{Conclusions and outlook}
\label{sec:conclusions}
In this paper, we have provided a calculation
of bare parton distribution functions using lattice QCD
techniques.  In particular, we analyzed the unpolarized, the
helicity and the transversity PDFs. In the cases of the
unpolarized and helicity PDFs, we have found a good
qualitative agreement with the phenomenologically extracted
PDFs.  In case of the transversity PDF, the uncertainties
from the phenomenological analyses are rather high such that the
lattice calculations --after a suitable renormalization--
have the potential to provide eventually the first theoretical
prediction which is, moreover, based only on QCD.  As a
general observation, the results of our ab initio,
non-perturbative lattice calculations show an asymmetry
between the quark and the anti-quark distributions, which is
a highly non-trivial outcome.

Employing the high statistics analysis in the case of the Gaussian smearing allowed us to reduce the
errors for the calculated matrix elements for $P_3=4\pi/L$ and
$6\pi/L$ by a factor of 2.5 as compared to our previous work
\cite{Alexandrou:2015rja}. This, in turn, leads to a much better
controlled matching to the physical PDFs, which is, however,
limited to a maximum momentum of $6\pi/L$.
A very promising new direction, to circumvent the problem of having
access to only low values of the nucleon momentum, is the use of momentum smearing
\cite{Bali:2016lva}.  We have tested this new smearing
technique and found large, $\mathcal{O}(10-100)$, factors of improvement
in the signal-to-noise ratio for our matrix elements, and could thus
perform the computation of the distributions for a momentum as large
as $10\pi/L$. This is an enormous step forward for the computation of
PDFs directly in lattice QCD, and Figs.~\ref{PDF_MS_UNPOL_M4M5} and \ref{FIG_PDF_MS_COMPARE_MOM_q0}
are our main results in this
respect. The resulting quark distributions at these large values of momentum
show the correct support in $x$ and, additionally, that their dependence on
$P_3$ starts to become weaker.

Although we only tested the momentum smearing technique on
a small number of gluon field configurations, and did not use
the full statistics available,  we are progressing to use this
technique on our ensembles at the physical
value of the pion mass
\cite{Abdel-Rehim:2015owa,Abdel-Rehim:2015pwa}, employing
there the full available statistics. Thanks to this new technique, we have now the
prospect to obtain accurate results for high momenta and
thus a well controlled matching to the physical PDFs.

Given the results of Fig.~\ref{PDF_MS_UNPOL_M4M5},
the remaining difference to the phenomenologically
extracted PDFs is, most probably, due to the large pion mass
and to the missing renormalization of the lattice PDFs.
The suspicion related to the renormalization is corroborated by
comparing results of non-smeared and HYP smeared lattice PDFs -- we
observe that the HYP smeared PDFs are much closer to the phenomenological ones.
However, it is also clear that a significant part of the difference between
our result and the phenomenologically extracted PDFs is due to the non-physical pion mass.
Computations of hadron structure observables from ensembles at the physical point
eliminated a large part of discrepancies with respect to the experimental values.
Such discrepancies were typically found in earlier studies that used larger than
physical pion masses \cite{Constantinou:2014tga,Abdel-Rehim:2015owa,Bali:2016wqg, Bhattacharya:2015wna, Durr:2015dna,Ohta:2014rfa, Green:2014xba, Green:2012ud, Lin:2014gaa}.
In particular, this concerns the first moment of the unpolarized iso-vector
PDF, $\langle x\rangle_{u-d}$, for which values in the ballpark between 0.2 and 0.3
were obtained at non-physical pion masses, while studies at the physical point
resulted in values close to the experimental one.
Hence, the plausible effect of decreasing the pion mass is to shift the curves like the ones
in Fig.~\ref{FIG_PDF_MS_COMPARE_MOM_q0} to the left (in the large-$x$ region), as this decreases
$\langle x\rangle_{u-d}$ obtained upon integration of these curves.
In the end, we expect that the computation at the physical pion mass, together with
renormalization and using large momenta, accessible with the momentum smearing technique,
will bring the lattice PDFs very close to the phenomenological ones.
We would like to note that since the submission of this work substantial progress has been made on the
renormalization of the quasi-PDFs. 
The authors of Ref.~\cite{Constantinou:2017sej} revealed a finite mixing pattern in lattice regularization for certain Dirac
structures~\cite{Constantinou:2017sej},
which led to the development of a complete non-perturbative
prescription~\cite{Alexandrou:2017huk}, which was followed closely by \cite{Chen:2017mzz}.
Nevertheless, all other possible systematic effects, like cut-off effects, will also need to be addressed for the ultimate comparison with phenomenology.

\section*{Acknowledgments} We thank our fellow members
of ETMC for their constant
collaboration. In particular helpful discussions with G.C.
Rossi are gratefully acknowledged.  We are grateful to the
John von Neumann Institute for Computing (NIC), the
J{\"u}lich Supercomputing Center and the DESY Zeuthen
Computing Center for their computing resources and support.
This work has been supported by the Cyprus Research
Promotion Foundation through the  Project Cy-Tera (NEA
Y$\Pi$O$\Delta$OMH/$\Sigma$TPATH/0308/31) co-financed by the
European Regional Development Fund. K.C.\ was supported in part by the
Deutsche Forschungsgemeinschaft (DFG), project nr. CI 236/1-1.
FS was partly supported by CNPq contract number 249168/2013-8. This work has
received funding from the European Union’s Horizon 2020 research and
innovation programme under the Marie Sklodowska-Curie grant agreement No 642069 (HPC-LEAP).

\bibliographystyle{utphys}
\bibliography{./references}

\end{document}